\documentclass[12pt,english]{article}
\usepackage[T1]{fontenc}
\usepackage[latin9]{inputenc}
\usepackage{geometry}
\usepackage{float}
\usepackage{textcomp}
\usepackage{url}
\usepackage{amsmath}
\usepackage{amssymb}
\usepackage{graphicx}
\usepackage{setspace}
\usepackage[authoryear]{natbib}
\usepackage{xcolor} 

\makeatletter

\providecommand{\tabularnewline}{\\}

\newcommand{\lyxaddress}[1]{
	\par {\raggedright #1
	\vspace{0em}
	\noindent\par}
}

\usepackage{ae,lmodern}
\usepackage{graphicx}
\usepackage{authblk}
\date{}

\renewenvironment{abstract}
 {\small
  \begin{center}
  \bfseries \abstractname\vspace{-.5em}\vspace{0pt}
  \end{center}
  \list{}{
    \setlength{\leftmargin}{.5cm}%
    \setlength{\rightmargin}{\leftmargin}%
  }%
  \item\relax}
 {\endlist}

\makeatother

\usepackage{babel}
\begin{document}

\title{\vspace{-1em} Seed banks can help to maintain the diversity of interacting phytoplankton
species}
\author{Coralie Picoche$^{1,2,*}$ \& Frédéric Barraquand$^{1,2}$}

\maketitle

\lyxaddress{\begin{center}
$^{1}$Institute of Mathematics of Bordeaux, University of Bordeaux
and CNRS, Talence, France\medskip{}
$^{2}$Integrative and Theoretical Ecology, LabEx COTE, University
of Bordeaux, Pessac, France
\par\end{center}}
$^{*}$corresponding author: \verb|coralie.picoche@u-bordeaux.fr|

\begin{abstract}
Seed formation is part of the reproductive cycle, leading to the accumulation
of resistance stages that can withstand harsh environmental conditions
for long periods of time. At the community level, multiple species
with such long-lasting life stages can be more likely to coexist.
While the implications of this process for biodiversity have been
studied in terrestrial plants, seed banks are usually neglected in
phytoplankton multispecies dynamic models, in spite of widespread
empirical evidence for such seed banks. In this study, we build a
metacommunity model of interacting phytoplankton species, including
a resting stage supplying the seed bank. The model is parameterized
with empirically-driven growth rate functions and field-based interaction
estimates, which include both facilitative and competitive interactions.
Exchanges between compartments (coastal pelagic cells, coastal resting
cells on the seabed, and open ocean pelagic cells) are controlled
by hydrodynamical parameters to which the sensitivity of the model
is assessed. We consider two models, i.e., with and without a saturating
effect of the interactions on the growth rates. Our results are consistent
between models, and show that a seed bank allows to maintain all species
in the community over 30 years. Indeed, a fraction of the species
are vulnerable to extinction at specific times within the year, but
this process is buffered by their survival in their resting stage.
We thus highlight the potential role of the seed bank in the recurrent
re-invasion of the coastal community, and of coastal environments
in re-seeding oceanic regions. Moreover, the seed bank enables populations
to tolerate stronger interactions within the community as well as
more severe changes to the environment, such as those predicted in
a climate change context. Our study therefore shows how resting
stages may help phytoplanktonic diversity maintenance.
\end{abstract}

Published in Journal of Theoretical Biology (2022) doi:\verb|10.1016/j.jtbi.2022.111020|
\clearpage{}

\section*{Introduction}

How the high biodiversity of primary producers maintains is still
an unresolved question for both experimental and theoretical ecology.
Terrestrial plants and phytoplanktonic communities can present hundreds
of species relying on similar resources, a situation where Gause's
principle implies that a handful of species should outcompete the
others. Some degree of niche differentiation, perhaps hidden to the
human observer, is generally expected for coexistence to maintain
\citep{chesson_mechanisms_2000}. However, a complex life-history
structure can further increase the likelihood of coexistence \citep[e.g.,][]{moll_competition_2008,fujiwara_coexistence_2011},
and so does the response of life history traits to variation in the
environment \citep{chesson_community_1988,huang_contribution_2016}.

Analyses of coexistence in terrestrial plant communities sometimes
take into account several life stages \citep[e.g.,][]{aikio_seed_2002,comita_asymmetric_2010,chu_large_2015},
though many models consider only a single life-stage \citep[see, among others,][]{ellner_alternate_1987,levine_effects_2004,martorell_testing_2014}.
When considering at least two stages, seeds/seedlings and adults,
several mechanisms that can contribute to long-term coexistence in
spatially and/or temporally fluctuating environment have been uncovered
\citep{shmida_coexistence_1985,chu_large_2015}.

The storage effect, a major paradigm in modern coexistence theory
\citep{chesson_mechanisms_2000,chesson_updates_2018}, which involves
a positive covariance between the strength of competition and favourable
environmental variation as well as buffered population growth, is
one of them. It has been shown to arise from a combination of interspecific
competition and a long-lived dormant stage, together with a temporally variable
recruitment rate \citep{caceres_temporal_1997}, which often arises
when recruitment is a direct function of environmental conditions
\citep{chesson_quantifying_2003}. The presence of a seed bank may
therefore lead to a storage effect \citep{angert_functional_2009},
though not systematically \citep{aikio_seed_2002}. However, although
the storage effect usually captures theoreticians' attention, the
contribution of seeds to coexistence may be much larger than their
potential contribution to the storage effect. A long-lived seed bank
can help coexistence by other, simpler means. For instance, in the
meta-community model of \citet{wisnoski_dormancy_2019}, when dormancy
and dispersal are present (without seed dispersal), local diversity
increases in temporally fluctuating environments. In their model,
adding a dormant stage could increase species diversity both at the
local and regional scales. These results suggest that considering
a seed stage in dynamical models can profoundly alter our understanding
of community persistence \citep[see also][]{jabot_macroecology_2017}.

Although there is some awareness of the role of cryptic life stages
in shaping terrestrial plant coexistence, the effect of such dormant
life stages on aquatic plant communities, and more specifically that
of phytoplanktonic algae, is often ignored. Such a gap is even more
surprising that phytoplankton organisms constitute one of the most
important photosynthetic groups on Earth, being responsible for half
the global primary production \citep{field1998primary}, and are the very basis of marine food
webs. The classical view behind phytoplankton dynamics is that their
blooms (peaks in abundances several orders of magnitude above their
baseline level) are due to seasonal variation in light, temperature
and nutrients, as well as hydrodynamic processes \citep{reynolds2006ecology}.
In this mindset, differential responses to environmental signals ensure
the coexistence of multiple species \citep{margalef_life-forms_1978,smayda_community_2001},
while always assuming that vegetative cells are already present in
the environment, or immigrating from a nearby water mass. Momentary
disappearances of a species are viewed as sampling issues at low density.
However, a complementary hypothesis suggests that resuspension and
germination of phytoplanktonic resting cells is another major player
allowing re-invasion from very low or locally zero population densities
\citep{patrick_factors_1948,marcus_minireview:_1998}. This long-standing
hypothesis is supported by recent reviews \citep{azanza_role_2018,ellegaard_long-term_2018}
which confirm that life history strategies including dormant individuals
are widespread in phytoplankton \citep[see][for extensive lists of species with known resting stages]{mcquoid_diatom_1996,tsukazaki_distribution_2018}.
Formation of a resting cell can either be part of the life cycle of
phytoplankton species and result from sexual reproduction or result
from asexual processes triggered by specific environmental conditions
\citep{ellegaard_long-term_2018}. Hereafter, when using the term
`seed banks', we refer to the accumulation of all types of resting
stages in the seabed, either sexual (dinoflagellate cysts)
or asexual (diatom spores or resting cells). A variety of models have endeavoured to
explain and predict amplitude, timing and/or spatial distribution
of blooms by explicitly modeling multiple stages in the life cycle
of a particular species, but without interactions with other organisms
(see for example \citealp{mcgillicuddy_mechanisms_2005,hense_towards_2006,hellweger_agent-based_2008,yniguez_investigating_2012}).
Two-to-four species \citep{yamamoto_modelling_2002,estrada_role_2010,lee_role_2018}
models also exist, but include a single species or compartement having
a resting stage. This state of affairs means that we currently have
no clear understanding of how the resting stage might help maintaining
biodiversity in species-rich communities. In the present paper, we
demonstrate the potential role of seed banks using a phytoplankton
community dynamics model.

Phytoplankton communities in coastal environments may benefit from
seed banks even more than the oceanic communities \citep[see for example][]{mcgillicuddy_mechanisms_2005},
as the distance to the sea bottom is smaller, which favours recolonization
from the sea bottom, something that is impossible in the deep ocean.
Moreover, `horizontal' exchanges between oceanic and coastal pelagic
phytoplanktonic communities are usually observed. A flow from the
ocean to coastal communities has been noticed for dinoflagellates
especially (\citealp{tester_gymnodinium_1997,batifoulier_distribution_2013}).
Conversely, in many other bloom-forming species, the shallower coastal
areas might function as a reservoir for biodiversity in the ocean.
Indeed, resting cells are able to germinate again after dozens of
years \citep{mcquoid_viability_2002,ellegaard_long-term_2018} or
even thousands of years \citep{sanyal2022not} of dormancy. Therefore,
we consider in this study three interlinked compartements: the coastal
pelagic environment, the seed bank, and the pelagic open ocean. The
coastal pelagic environment acts as a bridge between the seed bank
and the open ocean.

Our model is parameterized from field data (growth and interaction
rates within the phytoplankton community), and includes biotic and
abiotic constraints (e.g., particle sinking). In our analyses, we
examine how seed banks may influence the maintenance of biodiversity,
including under changing biotic interactions or changing environmental
conditions. We either add or remove the dormant compartment, which
allows to pinpoint its contribution to coexistence. We find that the
presence of resting stages prevents the extinction of several species.
Seed banks also allow a community to maintain its richness even with
strong disturbances of its interaction network, unless facilitative
interactions completely eclipse competitive interactions. Changes
in the environment, here represented by an increase in the mean temperature,
can also be buffered by seed banks. Finally, we discuss what information
would be required to further more accurate modeling of resting stages
in phytoplankton community dynamics.

\section*{Methods}

\subsection*{Models}

\begin{figure}[H]
\begin{centering}
\includegraphics[clip,width=0.99\linewidth]{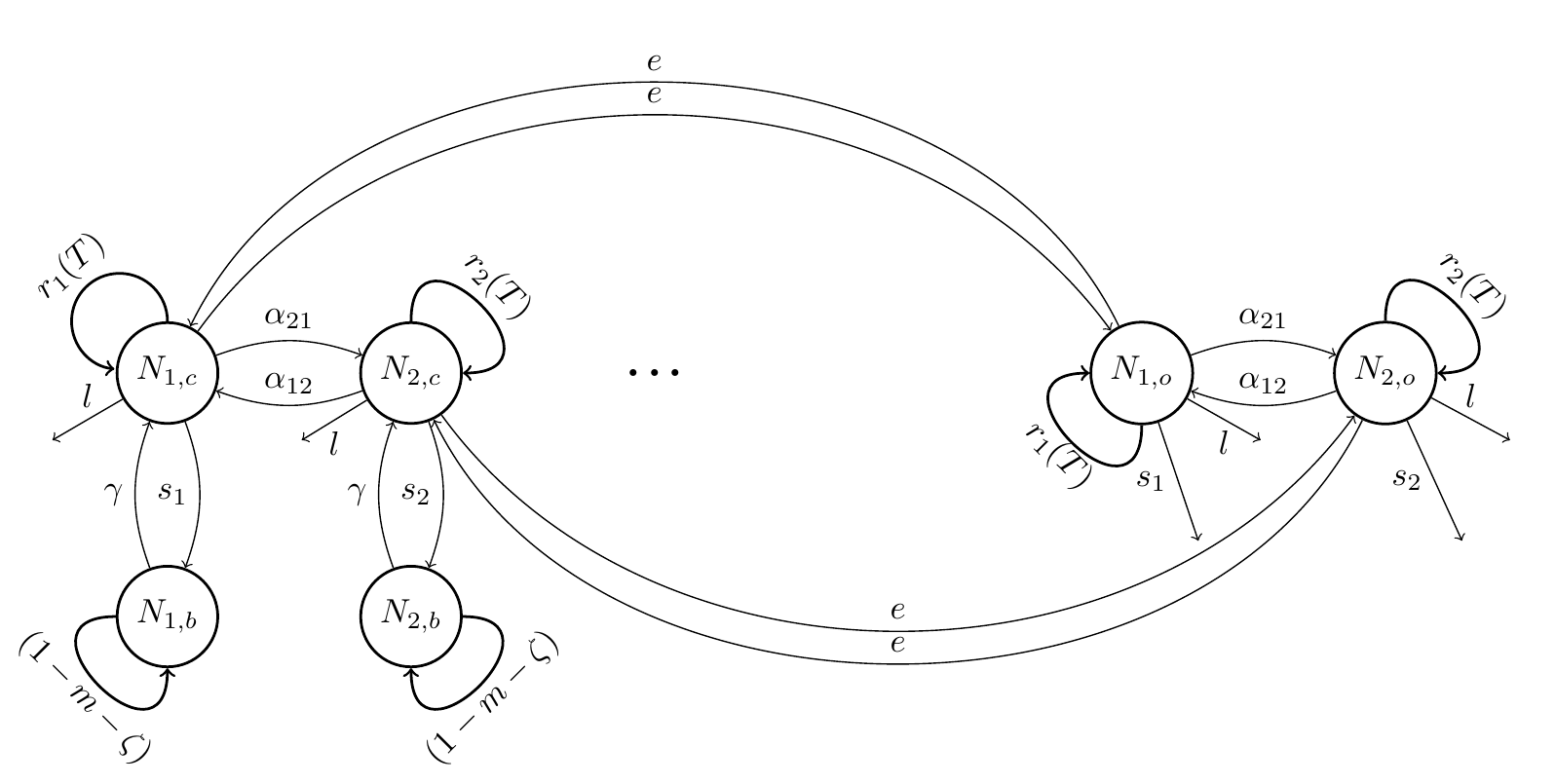}
\par\end{centering}
\caption{Structure of the model. Phytoplanktonic species (represented by circles)
are present in the coast (subscript c), the ocean (o) and the seed
bank (b). Parameters governing demography, interactions between organisms
and exchanges between compartments are defined in Table \ref{tab:State_Parameter}.
Only two species are shown here for the sake of simplicity but 11
species are present in the model. \label{fig:Conceptual-model-}}
\end{figure}

Our models (Fig. \ref{fig:Conceptual-model-}) build atop recent models
developed by \citet{shoemaker_linking_2016} and \citet{wisnoski_dormancy_2019},
although they diverge in several aspects developed below (e.g., possibility
for facilitative interactions). These discrete-time models are designed
for metacommunities with multiple interacting populations. Any discrete-time
model requires an ordering of events; in our models, these unfold
as follows: first, populations grow or decline according to a Beverton-Holt
(BH) multispecies density-dependence (eqs. \ref{eq:model_step1} and
\ref{eq:model_stepI_saturating}), and then, in a second step, exchanges
occur between the different compartments or patches constituting the
metacommunity (eq. \ref{eq:model_step2}).

In this paper, individuals are phytoplanktonic cells that move between
the upper layer of coastal water, its bottom layer where a seed bank
accumulates in the sediment, and the oceanic zone surrounding the
coastal water masses (hereafter ``the ocean\textquotedbl ). Only
oceanic and coastal pelagic cells are subject to BH-density dependence.
Resting cells in the seed bank are only affected by mortality $m$
and burial due to sedimentation $\zeta$. Parameters and state variables
are defined in Table \ref{tab:State_Parameter}.

\begin{table}[H]
\noindent\resizebox{0.99\textwidth}{!}{%
\begin{centering}
\begin{tabular}{|c|c|c|c|}
\hline 
Parameter & Name & Value (unit) & Status\tabularnewline
\hline 
$N_{t,i,c/o/b}$ & Abundance of species $i$ at time $t$ in the coast ($c$), ocean
($o$), or coastal bank ($b$) & NA (Number of cells) & Dynamic\tabularnewline
$T$ & Temperature & NA ($K$) & Dynamic\tabularnewline
$r_{i}(T)$ & Intrinsic growth rate of species $i$ & NA (day$^{-1}$) & Dynamic\tabularnewline
$b_{i}$ & Thermal decay of species $i$ & Field-based ($K^{3}$) & Calibrated\tabularnewline
$T_{i}^{opt}$ & Optimal temperature for species $i$ & Field-based ($K$) & Calibrated\tabularnewline
$d$ & Daylength & $50$ (\%) & Fixed\tabularnewline
$\alpha_{ij}$ & Interaction strength of species $j$ on $i$ in model I & Field-based (Cells$^{-1})$ & Calibrated\tabularnewline
$a_{C}/a_{F}$ & Maximum competitive/facilitative interaction strength in model II & Field-based (NA) & Calibrated\tabularnewline
$H_{ij}$ & Half-saturation for the interaction strength of species $j$ on $i$
in model II & Field-based (Cells$)$ & Calibrated\tabularnewline
$s_{i}$ & Sinking rate of species $i$'s resting cell & $\{0.1;\mathbf{0.3};0.5\}\times \text{Beta}(0.55,1.25)$ (day$^{-1}$) & Fixed\tabularnewline
$e$ & Exchange rate between ocean and coast & $\left\{ 0;\boldsymbol{0.4};0.9\right\} $ (day$^{-1}$) & Fixed\tabularnewline
$l$ & Loss rate of pelagic phytoplanktonic cells & $\left\{ 0.04;0.1;\mathbf{0.2}\right\} $ (day$^{-1}$) & Fixed\tabularnewline
$m$ & Resting cell mortality rate & $\mathbf{10^{-5}}$ (day$^{-1}$) & Fixed\tabularnewline
$\zeta$ & Resting cell burial rate & $\left\{ 10^{-3},\mathbf{10^{-2}},10^{-1}\right\} $ (day$^{-1}$) & Fixed\tabularnewline
$\gamma$ & Germination $(\gamma_{1})$ $\times$ Resuspension $(\gamma_{2})$
rate & $\left\{ 10^{-3},\mathbf{10^{-2}},10^{-1}\right\} \times\left\{ 10^{-5},10^{-3},\mathbf{10^{-1}}\right\} $
(day$^{-1}$) & Fixed\tabularnewline
\hline 
\end{tabular}
\par\end{centering}
}

\caption{Definition of main state variables and model parameters. State variables
and fluctuating parameters are indicated in the last column as ``Dynamic''.
Parameters that are constant through time are either ``Fixed'' (directly
obtained from literature) or ``Calibrated'' (obtained through model
fitting, with initial values arising from previous studies at the
study site). When a range of values is given, the bold numbers indicate
the reference values while the others are used for sensitivity analysis.
Beta(0.55,1.25) is the Beta distribution with parameters 0.55
and 1.25. For $\gamma,$ germination values for sensitivity analysis
were multiplied by the reference value for resuspension, and conversely.
\label{tab:State_Parameter}}
\end{table}

The Beverton-Holt (BH) formulation of multispecies population dynamics,
sometimes called Leslie-Gower \citep{cushing2004some}, is a Lotka-Volterra
competition equivalent for discrete-time models, and is often used
to represent terrestrial plant community dynamics. In our implementation
of the model, the population growth rate is modified by both
competitive and facilitative interactions, which translates into positive
and negative $\alpha_{ij}$ coefficients, respectively. We present
two different interaction models. We first use the classical multispecies
BH model (model I, eq. \ref{eq:model_step1}), also present in the
original models of \citet{shoemaker_linking_2016} and \citet{wisnoski_dormancy_2019}.
However, the high number of facilitative interactions characterizing
the modeled phytoplankton community \citep{picoche_strong_2020} combined
to the mass-action assumption could have very irrealistic destabilizing
consequences, which have been likened to an ``orgy of mutual benefaction''
\citep{may_1981_theoretical}: populations grow to infinity because
there is no saturation of beneficial effects when density increases.
In model I, we therefore forbid the realized growth rate to go above
the intrinsic growth rate (its theoretical limit), by setting a minimum
value of 1 to the denominator of the BH formulation. We subsequently
define saturating interactions, inspired by \citet{qian_balance_2020},
in our model II (eq. \ref{eq:model_stepI_saturating}). Model II provides
a more process-orientated solution to the issue of excessive mutual
benefaction (but at the cost of added parameters). Setting a miminum value of 1 to the denominator is still required for large increases or decreases in interaction strengths.

In our framework, the first step of model I can be written as

\begin{equation}
\left\{ \begin{array}{ccc}
N_{t',i,c} & = & \frac{\exp(r_{i}(T))N_{t,i,c}}{1+\sum_{j}\alpha_{ij}N_{t,j,c}}-lN_{t,i,c}\\
N_{t',i,o} & = & \frac{\exp(r_{i}(T))N_{t,i,o}}{1+\sum_{j}\alpha_{ij}N_{t,j,o}}-lN_{t,i,o}\\
N_{t',i,b} & = & N_{t,i,b}(1-m-\zeta)
\end{array}\right.\label{eq:model_step1}
\end{equation}

where the intrinsic growth rate $r_{i}(T)$ is a species-specific
function of the temperature (see eq. \ref{eq:growth_rate}), the interaction
coefficients $\alpha_{ij}$ are per capita effects of species $j$
on species $i$, and the loss term $l$ accounts for lethal processes
such as natural mortality, predation or parasitism. First estimates
of interaction coefficients are inferred from a previous study of
coastal community dynamics with Multivariate AutoRegressive (MAR)
models \citep{picoche_strong_2020}. We later calibrate these coefficients
for model I, since MAR models were applied at a different timescale.

The intrinsic growth rate $r_{i}(T)$ is defined through a modified
version of the formula used by \citet{scranton_coexistence_2016}
(eq. \ref{eq:growth_rate}), which classically decomposes $r_{i}(T)$
in two parts: the species-independent metabolism part $E(T)$ and
the species-specific niche part $f_{i}(T)$:

\begin{eqnarray}
r_{i}(T) & = & E(T)f_{i}(T)\label{eq:growth_rate}\\
\text{where }E(T) & = & d\times0.81e^{0.0631(T-273.15)}\nonumber \\
\textnormal{and }f_{i}(T) & = & \begin{cases}
\exp(-|T-T_{i}^{opt}|^{3}/b_{i}), & T\leq T_{i}^{opt}\\
\exp(-5|T-T_{i}^{opt}|^{3}/b_{i}), & T>T_{i}^{opt}.
\end{cases}\nonumber 
\end{eqnarray}

The metabolism part describes the maximum achievable intrinsic growth
rate based on \citet{bissinger_predicting_2008}. This maximum daily
intrinsic growth rate is weighted by the daylength $d$ as no growth
occurs at night. The realized niche part $f_{i}(T)$ describes the
decrease in growth rate due to the difference between the temperature
in the environment and the species-specific thermal optimum $T_{i}^{opt}$,
and is controlled by the species-specific thermal decay $b_{i}$,
which depends on the niche width. It is important to note that unlike
$E(T)$, which models direct effects of temperature on metabolism,
$f_{i}(T)$ is a phenomenological construct including all indirect
effects of temperature, mediated by environmental variables correlated
to temperature (light, nutrient inputs, predation, among other environmental
conditions). In other words, $f_{i}(T)$ corresponds to a realized
niche, pertaining to a given environment, which can be much more narrow
than the fundamental thermal niche. Parameterisation is further detailed
in Section S1 of the SI.

In model II, oceanic and coastal dynamics are governed by eq. \ref{eq:model_stepI_saturating}:

\begin{equation}
N_{t',i,c/o}=\frac{\exp(r_{i}(T))N_{t,i,c/o}}{1+\sum_{j\in\mathbb{C}}\frac{a_{C}N_{t,j,c/o}}{H_{ij}+N_{t,j,c/o}}+\sum_{j\in\mathbb{F}}\frac{a_{F}N_{t,j,c/o}}{H_{ij}+N_{t,j,c/o}}}-lN_{t,i,c/o}\label{eq:model_stepI_saturating}
\end{equation}

where $a_{C}$ and $a_{F}$ are the maximum competition and facilitation
strengths, respectively, with $\mathbb{C}$ and $\mathbb{F}$ the
sets of competitors and facilitators of species $i$. We use here
similar notations to \citet{qian_balance_2020}, but use different
parameters that vary between species. Indeed, the half-saturation
coefficients $H_{ij}$ vary between species, as opposed to the maximum
rates in \citet{qian_balance_2020}. It did not make sense biologically
for $H_{ij}$ to be fixed (e.g., in a resource competition context,
different species are expected to feel resource limitations at different
concentrations of nutrients and at different numbers of competitors).
How to shift from MAR- to BH-interaction matrices in model I, and
to use the parameter estimates of model I to specify parameters in
model II is described in Section S2 of the SI.

After growth and mortality processes occur, exchanges take place between
the three compartments, which constitutes the second step of the model
(eq. \ref{eq:model_step2}):

\begin{equation}
\left\{ \begin{array}{ccc}
N_{t+1,i,c} & = & (1-s_{i}-e)N_{t',i,c}+\gamma N_{t',i,b}+eN_{t',i,o}\\
N_{t+1,i,o} & = & (1-s_{i}-e)N_{t',i,o}+eN_{t',i,c}\\
N_{t+1,i,b} & = & (1-\gamma)N_{t',i,b}+s_{i}N_{t',i,c}
\end{array}\right.\label{eq:model_step2}
\end{equation}

Each compartment (ocean, coast, coastal seed bank) contains $10^{3}$
cells at the beginning of the simulation, and the dynamics are run
for 30 years with a daily time step. We model the temperature input
as a noisy sinusoidal signal with the same mean and variance as the
empirical data set described below: the amplitude of the sinusoid
is 12.4°C and the standard deviation of the noise is 0.25°C.

\subsection*{Parameterization of the models}

\subsubsection*{Literature-derived parameter values}

\paragraph*{Loss rate}

The loss rate $l$ of vegetative cells can be attributed to natural mortality,
predation or parasitism. This rate is quite variable in the literature:
the model of \citet{scranton_coexistence_2016} considered a rate
around 0.04 day$^{-1}$ while a review by \citet{sarthou_growth_2005}
indicates a grazing rate of the standing stock between 0.2 and 1.8
day$^{-1}$ and an autolysis rate between 0.005 and 0.24 day$^{-1}$
(in the absence of nutrients, or because of viral charge). A maximum
value of 0.2 is fixed for the model, as a balance between using a
high loss rate (probable because of predation) and keeping all species
in the community in the reference model (see Section S3 of the SI
for more details).

\paragraph*{Sinking rate}

Phytoplanktonic particles have a higher density than water and cannot
swim to prevent sinking (although they are able to regulate their
buoyancy, \citealt{reynolds2006ecology}). Sinking is mostly affected
by hydrodynamics, but at the species-level, size, shape, density-regulation
and colony-formation capabilities are key determinants of the particle
floatation. In this model, the sinking rate of each species resting
cells is drawn from a Beta distribution with a mean value of 9\%,
and a maximum ($S_{\text{max}}$) around 30\%, that is $s\backsim0.3\beta(0.55,1.25)$
(see Fig. S4), adapted from observations by \citet{passow_species-specific_1991}
and \citet{wiedmann_seasonality_2016}.

\paragraph*{Exchange rate}

The exchange rate $e$ between the ocean and the coast depends on the shape
and location of the coast (estuary, cape, ...). Our calibration site
is located at an inlet. The flow at the inlet leads to a renewal time
of the coastal area water evaluated between 1 and 2.5 days \citep{ascione_kenov_water_2015},
which corresponds to a daily exchange rate between 40 and 100 \%.

\paragraph*{Mortality and burial in the seed bank}

Loss from the seed bank is the result of cells' mortality $m$ and
their burial by sedimentation $\zeta$. Mortality values range between
$10^{-5}$ and $10^{-4}$ per day (more details on the approximation
of mortality rates from \citealt{mcquoid_viability_2002} are given
in Section S3 of the SI). However, burial by sedimentation is the
prevailing phenomenon. Indeed, once resting cells have been buried,
they are not accessible for resuspension even if they could still
germinate. Burial depends on the hydrodynamics of the site, but also
on biotic processes (i.e., bioturbation) and anthropogenic disturbances
such as fishing or leisure activities (e.g., jet skiing). This parameter
is thus heavily dependent on the environmental context and varies
here between 0.001 and 0.1 per day.

\paragraph*{Germination/resuspension}

Both germination ($\gamma_{1})$ and resuspension ($\gamma_{2}$)
are needed for resting cells to contribute to the vegetative pool
in the water column ($\gamma=\text{resuspension\ensuremath{\times}germination})$.
As actual rates of germination are not easily deduced from the literature,
a set of credible values has been tested (1\%, 0.1\%, 0.01\%). Similarly,
resuspension values are seldom computed for phytoplanktonic cells,
but models for inorganic particles can be used (see Section S3 of
the SI for literature and details). In this paper, we explore values
between $10^{-5}$ (stratified water column) to 0.1 (highly mixed
environment).

\subsubsection*{Initial interaction matrix}

Initial values of interaction strengths between species are based
on Multivariate AutoRegressive (MAR) model estimates \citep{picoche_strong_2020}.
MAR(1) models relate the log-abundance of each of the $S$ phytoplankton
species at time $t+1$ to log-abundances of all species at time $t$,
through an interaction matrix, and effects of abiotic variables at
time $t+1$ (see Section S2 of the SI). Interactions are estimated
only between species within the same trophic level, and are independent
from the environmental variables that were included in the MAR model
estimates as covariates, such as temperature in our case. This allows
to remove at least some of the confounding factors, such as seasonality.
A phylogeny-based interaction matrix resulted in a better fit to model
the community dynamics, i.e. pennate/centric diatoms only interact
with other pennate/centric diatoms, respectively, and dinoflagellates
only interact with dinoflagellates.

The MAR model can only estimate apparent interaction strengths: complex
processes can be at work behind values of competition and facilitation,
either from abiotic (e.g., hydrodynamics) or biotic (e.g. consumption
by predators or parasites) variables. For more consideration on apparent
interactions detected in phytoplankton communities with the MAR model,
we refer the reader to \citet{barraquand_coastal_2018}. In the present
model, while we tune interaction strength values (see below), the
type of interaction (competition, facilitation or absence of interaction)
remains the same as computed in \citet{picoche_strong_2020}.

\smallskip{}

\subsubsection*{Parameter calibration}

As explained above, we use initial interaction estimates from our
previous time series modelling \citep[see Section S3 of the SI for the equations]{picoche_strong_2020},
which are then calibrated to the time series (thus possibly re-estimated,
see below), to take into account the differences in model structure
and timescale between this study and \citet{picoche_strong_2020}.
In SI Section S3, we present the formulas relating the MAR interaction
coefficients to the Jacobian matrices of the Beverton-Holt multispecies
models, as these formulas allow to obtain proper $\alpha_{ij}$ coefficient
values.

The calibration procedure consists in launching 1000 simulations,
each characterized by a specific set of interaction coefficients.
More precisely, for each simulation, an interaction coefficient ($\alpha_{ij}$
in model I, $H_{ij}$ in model II) has probability $1/5$ to keep
its present value, probability $1/5$ to increase by 10\% , $1/5$
to decrease by 10\%, $1/5$ to be halved and $1/5$ to be doubled.
The numbers of coastal pelagic cells (which are the ones measured
empirically) are then extracted over the last 2 years of the simulation,
and compared to observations using the following summary statistics:
\begin{itemize}
\item average abundance $f_{1}=\sqrt{\frac{1}{S}\sum_{i}^{S}\left(\bar{n}_{i,obs}-\bar{n}_{i,sim}\right){}^{2}}$
where $S$ is the number of taxa and $\bar{n_{i}}$ is the logarithm
of the mean abundance of taxon $i$.
\item amplitude of the cycles $f_{2}=\sqrt{\frac{1}{S}\sum_{i}^{S}\left[\left(\max(n_{i,obs})-\min(n_{i,obs})\right)-\left(\max(n_{i,sim})-\min(n_{i,sim})\right)\right]{}^{2}}$
where $n_{i}$ is the logarithm of the abundance of taxon $i$.
\item period of the bloom. The year is divided in 3 periods, i.e. summer,
winter and the spring/autumn group (as taxa blooming in these periods
can appear in either or both seasons). We give a score of 0 if the
simulated taxon blooms in the same period as its observed counterpart and 1
otherwise.
\end{itemize}
Simulations with taxon extinction (i.e., the taxon is absent for more
than 6 months in a compartment) are discarded, as extinctions are
not observed in the field data. Parameter sets are then ranked according
to their performance for each summary statistic, and we select the
set of interactions minimizing the sum of the ranks.

\subsubsection*{Sensitivity analysis}

Parameters taken from the literature may be site- or model- specific,
or vary over several orders of magnitude in the literature, e.g.,
rates of sinking $s$, resuspension/germination $\gamma$, seed burial
$\zeta$, and loss of pelagic cells $l$. We therefore performed
a sensitivity analysis to these highly uncertain parameters. The set
of tested values for each parameter is given in Table \ref{tab:State_Parameter}.
We used average abundances and amplitudes at the community and taxon
levels for the last 2 years of simulations as the major model diagnostics.

\subsubsection*{Empirical dataset used for calibration}

The models are calibrated using time series of phytoplanktonic abundances
that have been monitored biweekly for 21 years in the Marennes-Oléron
Bay, on the French Atlantic Coast \citep[the Auger site analysed in][]{picoche_strong_2020}.
We stress that we are not trying to model precisely this particular
community, but rather to constrain our models with an empirically-derived
interaction network and species-specific thermal niches, which helps
to produce lifelike patterns of phytoplankton community dynamics resembling
observable data (seasonal dynamics, high-amplitude blooms, differences
in average abundances matching data). 

\subsection*{Scenarii}

The effect of the seed bank on biodiversity and community dynamics
can be evaluated through the response to disturbance with and without
the resting-stage compartment. Mortality in the seed bank is set to
100\% to effectively remove the compartment. We evaluate two main
disturbances:
\begin{enumerate}
\item increase or decrease in interaction strength
\item temperature change, either in mean value or variability
\end{enumerate}
In the first scenario, interaction strengths are multiplied or divided
by a factor ranging between 1 and 10. In order to differentiate the
effects of facilitative and competitive interactions on coexistence,
we vary only one type of interactions at a time. Here, both intra
and interspecies interactions are modified; we present in Section
S6 of the SI additional simulations with a change in interspecies
interactions only.

In the second scenario, five different climate change trajectories
are assessed. In the first three, the average temperature is increased
by 2, 5, or 7°C \citep{boucher_presentation_2020}. In the next two,
keeping the reference average temperature, the total variance of the
temperature, including seasonality and noise, is either decreased
or increased by 25\%. Each climate change trajectory is run 5 times
to account for the intrinsic stochasticity of the temperature signal.

In both scenarii, simulations are run for 30 years for both population
growth models, with and without a seed compartment, and only the last
2 years are considered to evaluate effects of changes in parameters
and in temperature. The code for all simulations is to be found at \url{https://github.com/CoraliePicoche/SeedBank}.

\section*{Results}

\subsection*{Phytoplankton dynamics}

\begin{figure}[H]
\begin{centering}
\includegraphics[height=0.8\textheight]{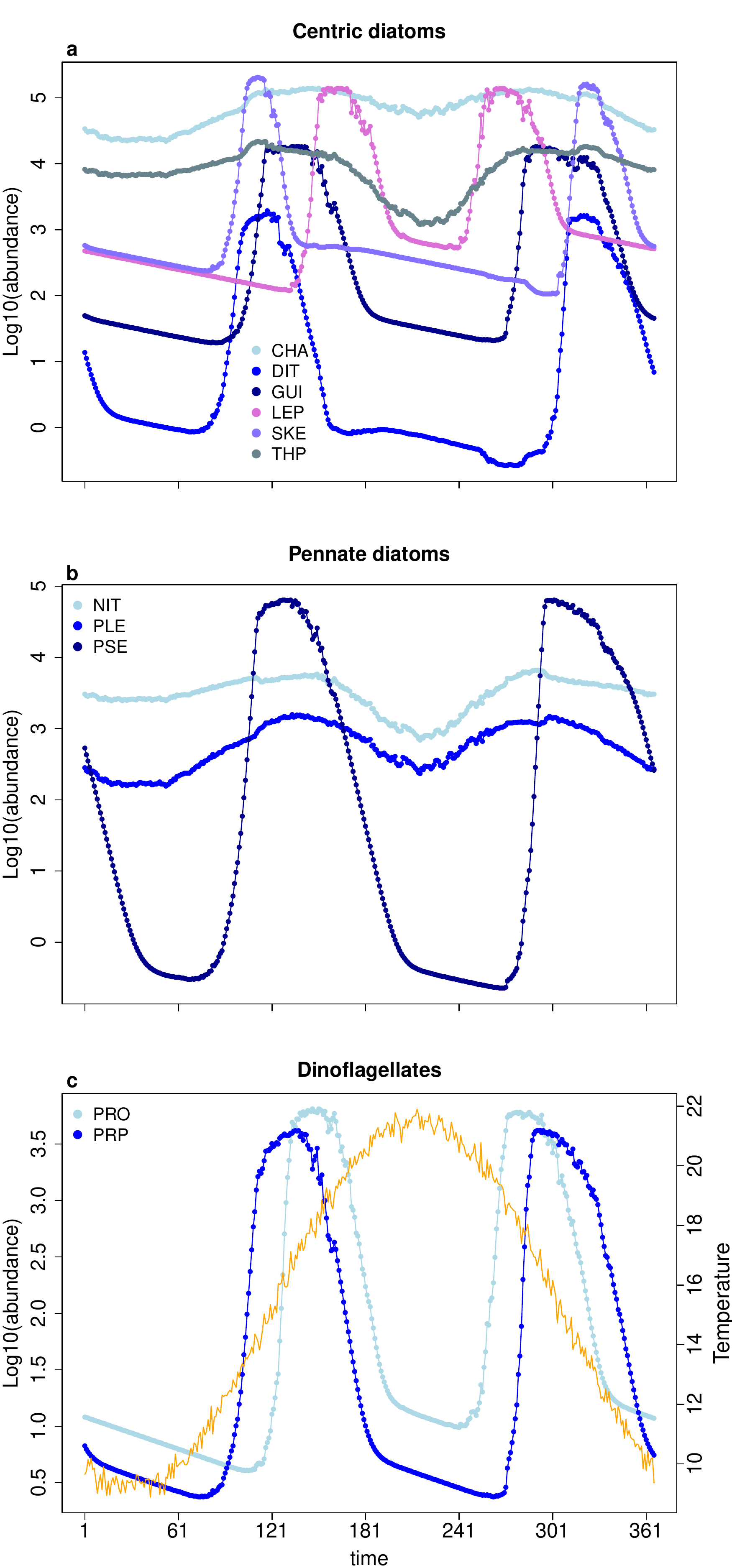}
\par\end{centering}
\caption{Simulated phytoplankton dynamics for a year in model I (time in days). Each panel
corresponds to a cluster of interacting taxa: centric diatoms (a),
pennate diatoms (b) and dinoflagellates (c). Taxa only interact within
their cluster of related taxa (see Methods). The orange line in the
third panel indicates the temperature.\label{fig:-Simulated-phytoplankton}}
\end{figure}

The classical mass-action (model I) and saturating interaction (model
II) formulations of multispecies dynamics both reproduced the main
characteristics of observed phytoplankton dynamics. They produced
one or two blooms during the year and a range of abundances covering
several orders of magnitude, with the right timing of the blooms.
At the Auger site that was used for calibration, abundances increase
in spring and can last over part of summer, or start a new bloom in
autumn, which is what we observed as well in the models. Annual mean
abundance of the various species was also well reproduced. That said,
in some cases, abundances could be lower than expected and the variation
in abundances due to seasonality was underestimated (Fig. \ref{fig:-Simulated-phytoplankton}).
In all cases, saturating interactions led to higher abundances than
mass-action interactions throughout the year (Fig. S5).

\subsection*{Sensitivity to uncalibrated parameters}

\begin{figure}[H]
\begin{centering}
\includegraphics[width=0.9\textwidth]{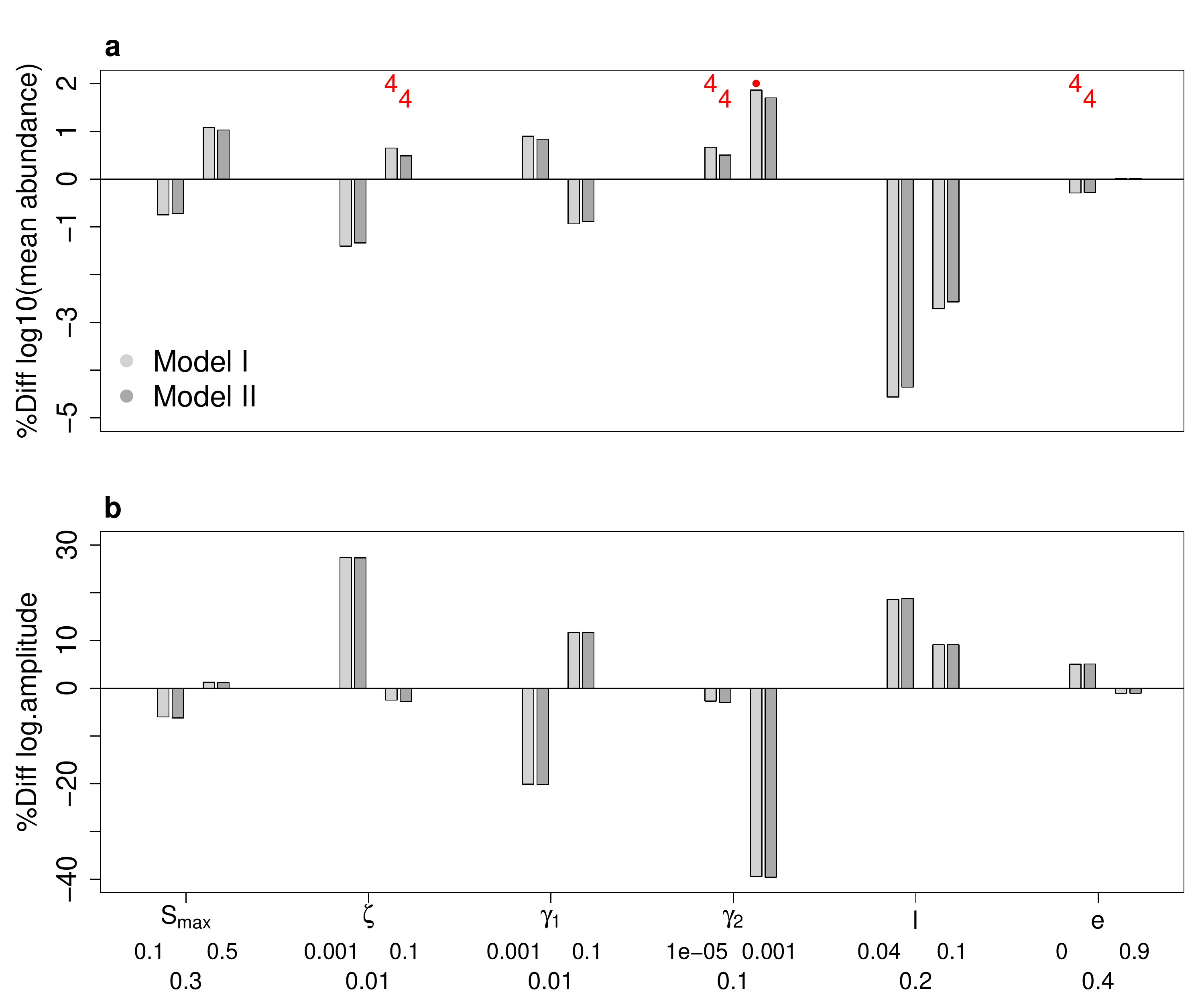}
\par\end{centering}
\caption{Sensitivity of the model to variation in parameters, measured as the
difference between the reference simulation metric and the metric
for the simulation including a change in parameter. The two metrics
used were the decimal log average abundance (a) and the decimal logarithm
of the ratio between maximum and minimum abundance (b) of the coastal
phytoplanktonic community. The x-axis is divided in three rows: the
symbol of the parameter defined in Table 1 (first row), values used
in the sensitivity analysis (second row) and values used in the reference
simulation (third row). Numbers in red in the top panel are the final
number of species in the ocean and dots correspond to simulations
in which at least one species abundance reached 0 at one point but
the species did not disappear. \label{fig:Sensitivity}}
\end{figure}

Phytoplankton abundances were not strongly affected by changes in the
parameter values (Fig. \ref{fig:Sensitivity}). As parameters were
varied in their plausible range, the average change in mean abundance
on the coast between the reference simulation and the sensitivity
simulations varied between -4.6 and 1.9\% for model I and between
-4.4 and 1.7\% for model II, with similar deviations (same sign and
magnitude) in the two models. 

In the two models, the decrease in mortality rate of vegetative cells
$l$ had the highest impact on the final average abundance, leading
to an increase in abundances. The exchange rate between the ocean
and the coast had a much lower effect on the coastal average abundance.

On the other hand, the amplitude (i.e., decimal logarithm of the maximum
to minimum ratio of abundance) was more affected by changes in parameters
and could vary by -39.6 to 18.8\% in model I, and between -39.6\%
and 18.8\% in model II. Results were qualitatively the same in the two
models, with a decrease in resting-stage burial being the main driver
of the decrease in amplitude, and a decrease in resuspension leading
to an increase in amplitude.

In three cases (burial rate set to 0.1, resuspension set to 10$^{-5}$
or exchange rate set to 0), the final richness of the oceanic community
decreased from 11 to 4. Extant species were the same in all simulations
(CHA, THP, NIT, PLE, i.e. temperature generalist species; the correspondence
between codes and groups of species is given in Section S4 of the
SI). When resuspension was set to 0.001, a species periodically disappeared
from the ocean, to be subsequently re-seeded by the coastal population.

For all parameters, except the sinking rate, an increase in mean abundance
was associated to a decrease in amplitude.

\subsection*{Scenarii of environmental change}

Two scenarii were designed to test the buffering effect of the seed
bank against disruption. In both cases, it consisted in removing the
seed bank by setting resting-stage mortality to 100\% per day. Without
any other disturbance to the system, this led to a decrease in richness
from 11 to 4 species at the end of the simulation (Fig. \ref{fig:Seed_bank_compet})
while the total abundance of phytoplankton was not strongly affected
(around $10^{5}$ in all cases). The inverse of the Simpson index
(the second Hill number) decreased from approximately 3 to 1, showing
that the disappearance of the seed bank did not affect only the rarest
species.

\subsubsection*{Biotic effects}

\begin{figure}[H]
\begin{centering}
\includegraphics[width=0.7\textwidth]{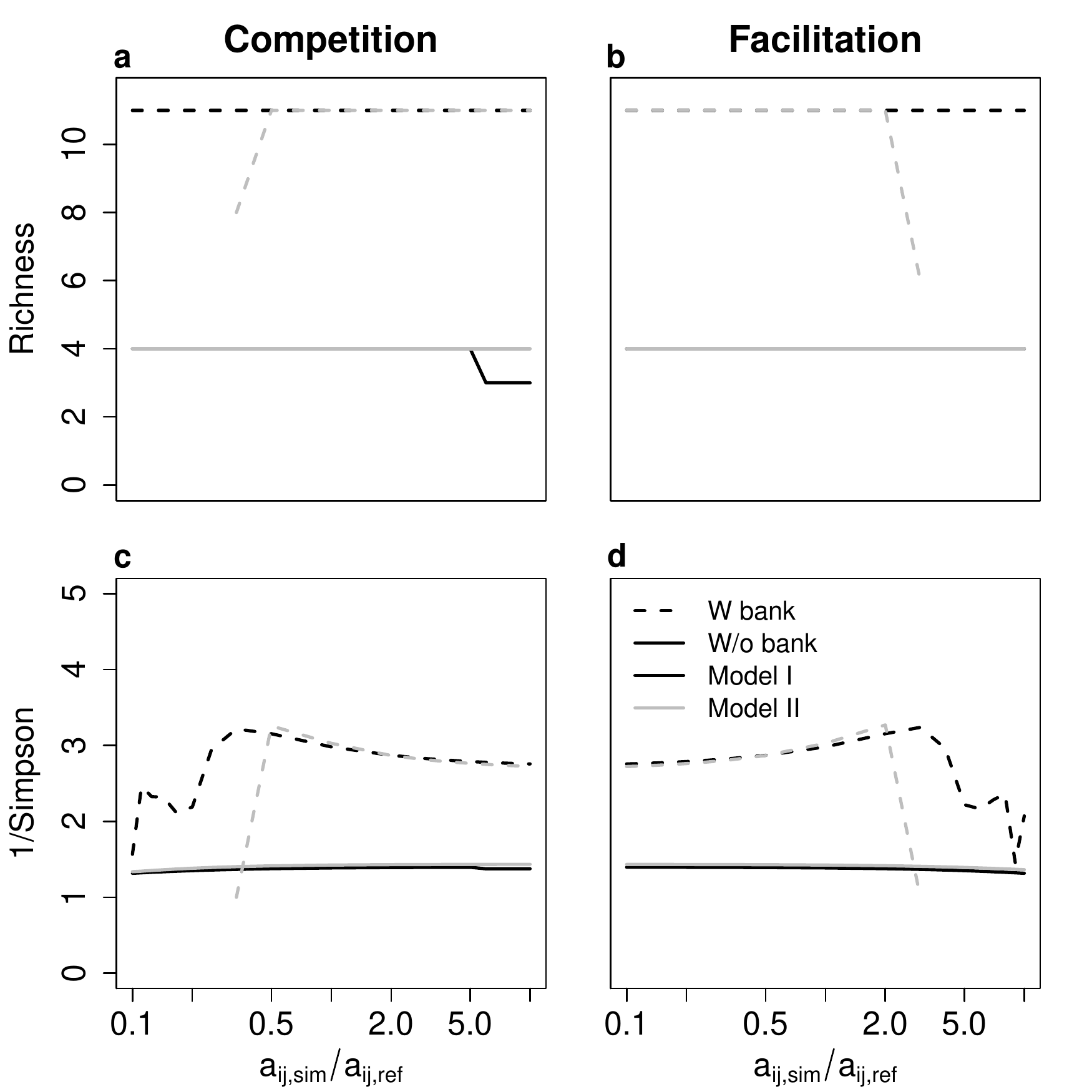}
\par\end{centering}
\caption{Measures of biodiversity in the ocean at the end of the simulation:
a-b) richness and c-d) inverse of the Simpson index, with (dashed
line) and without (solid line) a seed bank, as a function of the strength
of competition and facilitation with a classical Beverton-Holt (black
lines) or a saturating interaction (grey lines) formulation. The x-axis
shows the factor by which each interaction coefficient was multiplied, e.g. the
value 0.1 indicates that the interaction strengths in the simulation
are 10 times lower than the interaction strengths in the reference
simulation. Note the logarithmic scale.\label{fig:Seed_bank_compet}}
\end{figure}

Our first hypothesis was that the absence of the seed bank would cause
the community to be more affected by a higher competition strength.
Counter-intuitively, our results (Fig. \ref{fig:Seed_bank_compet})
showed that an increase in competition strength only had negative
effects with model I, and for high competition values (6 times the
reference ones at least), shifting from 4 to 3 species in the oceanic
compartment of a community without a seed bank. By contrast, an increase
in competition strength did not affect the richness of a community
with a seed bank. On the contrary, a decrease in competition (from
a factor 0.5 and lower) or an increase in facilitation (starting from
a factor 2 and higher) led to much smaller communities in model II
in the presence of a seed bank.

The inverse of the Simpson index was also affected by the changes
in interaction strengths, with similar patterns to richness, as it
was lowest for high facilitation or low competition. Some
species reached very high growth rates in these scenarios, which then
fed back onto community dynamics, generating lower diversity in the
end.

\begin{figure}[H]
\begin{centering}
\includegraphics[width=0.99\textwidth]{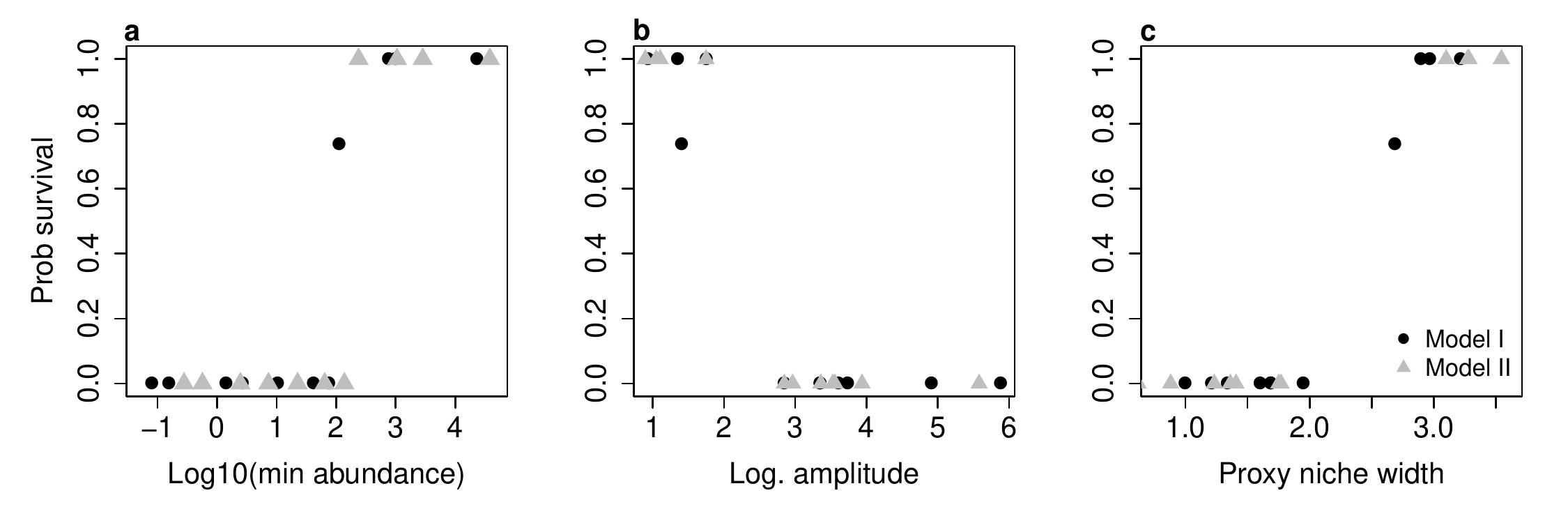}
\par\end{centering}
\caption{Probability of survival of species along a gradient of competition strength and in the
absence of a seed bank, as a function of their population dynamics characteristics
(minimum abundance, logarithm of amplitude and niche width) in the
reference parameter set.\label{fig:Probability-of-survival} Each
dot corresponds to a species, and its survival using model I (black)
or model II (grey).}
\end{figure}

Species which disappeared were characterized by a lower minimum abundance,
a higher amplitude of fluctuations and a small niche (Fig. \ref{fig:Probability-of-survival}).
However, their interactions were not qualitatively different from
the other species.

\subsubsection*{Abiotic effects}

\begin{figure}[H]
\begin{centering}
\includegraphics[width=0.7\textwidth]{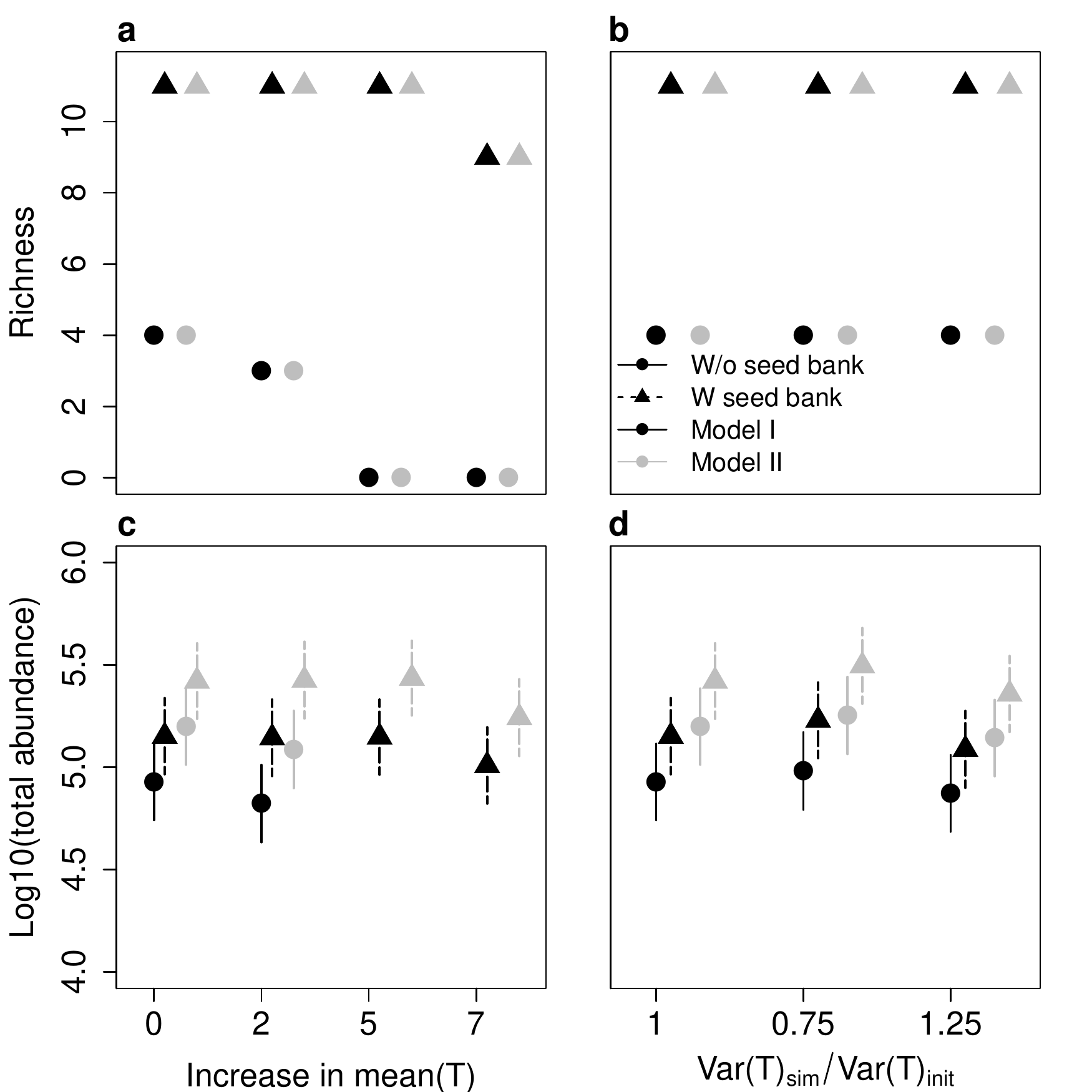}
\par\end{centering}
\caption{Variation in richness and total abundance with and without a seed
bank as a function of the mean (left) and variance (right) of the
temperature with a mass-action (black line, model I) or a saturating
interaction (grey line, model II) formulation. \label{fig:Seed_bank_temp}}
\end{figure}

Our second hypothesis was that the absence of a seed bank would reduce
the ability of a community to withstand changes in its abiotic environment,
here represented by variation in the temperature. This was true for
both models (Fig. \ref{fig:Seed_bank_temp}), as the communities without
a seed bank could not maintain their richness with an increase in
temperature above 2°C, as opposed to communities with a seed bank,
which could only be affected by a 7°C increase (scenario SSP5 8.5,
\citealt{boucher_presentation_2020}). In all cases however, the total
abundances were not strongly affected. Indeed, the total abundance
of a community is driven by a small number of numerically dominant
species, which did not disappear. High total abundances tended to
correspond to the abundance of only one or two species. Model II consistenly
led to higher abundances, as was already the case in the reference
simulations.

The variance of the temperature did not affect richness nor total
abundance of communities with a seed bank. This was also true without
a seed bank. The presence of resting stages did increase total abundance
though.

Some additional simulations were made to better understand the functioning
of the model, during which the temperature was kept constant (equal
to the average temperature of the fluctuating environment). For a
constant and average temperature, the presence of a seed bank did
not change the final richness of the communities. There
were 9 extant species at the end of the constant-environment simulations,
whether there was a seed bank or not. The two specialist species whose
thermal optima were farthest from the constant temperature went extinct
even when the seed bank was present, since the environmental fluctuations
allowing them to benefit from cold or warm temperature were removed.
The other species benefited from the constant environment, as 9 species
instead of 4 persisted in the case where the seed bank was removed
(see below for a discussion). The high persistence in a constant environment
(9 out of 11) translates the stabilizing, data-driven
interaction structure of our model. For both models I and II, the
dynamical system reached a fixed point equilibrium in absence of the
fluctuations induced by a variable temperature. 

\section*{Discussion}

Using a meta-community model which accounts for exchanges between
the ocean and the coast, as well as movements between the top and
the bottom of the coastal water column, we were able to show that
a resting stage, leading to the formation of a seed bank, can help
maintain biodiversity. Our phytoplanktonic
community dynamics model was parameterized based on literature,
field-based phenology, and interaction strength estimates. It
was then calibrated on phytoplankton community time series. Our model
was able to simulate realistic community dynamics (both mean abundances
and temporal patterns), while including the effects of both positive
and negative interactions on community dynamics. When removing the
seed bank, biodiversity decreased drastically. The total abundance
of the community decreased as well. This was true for the reference
parameter values, as well as when species interaction strengths and
environmental fluctuation levels were altered, in which cases the
seed bank's buffering influence disappeared. Moreover, when faced
with a biotic or abiotic perturbation, communities where
species could divert part of their population to a dormant stage were
less prone to species loss and could maintain their biomass through
the years. These results were consistent for the two interaction models
that we considered, with and without progressive saturation in interaction
strengths. Our results therefore demonstrate the major potential
role of phytoplanktonic resting stages in maintaining biodiversity.
These results align with the findings of previous theoretical studies,
that have put forward similar effects of dormant stages in other taxa,
such as plants \citep{levine_effects_2004,jabot_macroecology_2017},
invertebrates \citep{wisnoski_dormancy_2019} or (smaller) microbes
\citep{jones_dormancy_2010}.

The positive effect of a seed bank on species diversity in our model
is contingent upon environmental fluctuations. Indeed, in a constant
environment, the absence of a seed bank did not alter persistence,
and final richness was higher than in a fluctuating environment (9
species in a constant environment vs. 4 species in a variable environment,
both without a seed bank). The absence of abiotic perturbations of
the intrinsic growth rates enabled species which were not too far
from their thermal optima to maintain. This might be surprising for
readers acquainted with Hutchinson's nonequilibrium theory \citep{hutchinson_paradox_1961}.
However, the theoretical literature on persistence in variable environments
has shown that additional mechanisms \citep[reviewed in][]{fox_intermediate_2013}
are needed to maintain diversity in the long run, such as relative
nonlinearities or the storage effect, which is at least absent from
the pelagic part of our model due to the absence of buffered growth
(see Section S8 in SI). There is a clear destabilizing effect of environmental
variation in our model without the seed bank, which is probably heightened
by the fact that environmental variation is positively autocorrelated
through seasonality \citep{picoche_how_2019,schreiber_positively_2021}.
Although the storage effect could manifest itself when combining pelagic
and seed bank parts of our model, it is very unlikely to be at play
here as environmental conditions do not impact recruitment rates in
our model, while this is usually a requirement for the storage effect
to manifest in such models \citep{chesson_quantifying_2003}. 

Even in the absence of a storage effect, however, the high longevity
of the resting stage itself can explain the effect of the seed bank
in a fluctuating environment. Such longevity is due to dormancy, which
has long been observed in field and experimental data, including for
phytoplanktonic organisms \citep{eilertsen_phytoplankton_2000}, and
has been theorized to be an important and neglected process in the
wider microbiology literature \citep{locey_synthesizing_2010,jones_dormancy_2010}.
It allows recolonization of a community where counts of pelagic cells
alone would suggest that some species have gone extinct. This colonization-in-time
may of course combine with present recolonization from other spatial
areas, as is known for plants \citep{shmida_coexistence_1985}. In
our case, our focus on phytoplankton led us to assume that organisms
moved between the coast and the ocean, which were largely synchronous
environments. Spatial recolonization was therefore less important
than temporal recolonization; the relative importance of the two processes
may vary depending on the organisms and the degree of spatial synchrony
of their environment (in plankton, see \citealt{montresor_dinoflagellate_1998,anderson_alexandrium_2005}).

The specificities of phytoplankton resting stages, that usually fall
to the ocean bottom in coastal areas, led us to assume that only the
``vegetative'' stage (here, the classic pelagic form of planktonic
cells) disperse. In some other metacommunity models with dormant seed
banks \citep[e.g.][]{wisnoski_dormancy_2019}, the dormant stage can
disperse as well. This would be true for most plants too (and perhaps
some phytoplankters in situations where they are transported by animals).
However, the restriction about which stage can move did not change
the general conclusion already stated by \citet{wisnoski_dormancy_2019}:
the combination of spatial dispersal and dormancy through seed banks
greatly helps biodiversity maintenance. In our study, this main result
was also robust to changes in exchange parameters and mean interaction
values in the community.

Species persistence varied between the coastal and oceanic compartment.
Temporary species disappearance could happen periodically in the ocean
but species were able to reinvade from the coast, thanks to the connection
of the coast to the seed bank. Local extinction in the absence of
a seed bank confirms conclusions from \citet{hellweger_agent-based_2008}
on a single species. This suggests that some species may be locally
transient: they are filtered out from certain patches, but can reinvade
more or less periodically the environment (as found by \citealt{guittar_quantifying_2020}, for grasslands).

Survival probabilities of species within the community in the absence
of a seed bank also depended on their characteristics -- and therefore
species rescued by the seed bank also have specific characteristics.
While extinct species growth rates tended to be less affected by interactions for average densities (Fig. S7 in SI), such species had higher amplitudes of population variation as well as a smaller niche width. Specialists species, growing at specific temperatures, therefore typically benefit the most from a seed bank.

Despite the evidence for seed bank effects that we and others uncovered,
phytoplanktonic community models designed to explain biodiversity
usually avoid modelling seed banks. In our view, this may decrease
the possibility of spontaneous re-colonization at the coast (at very
low densities initially), which can then spill to the open ocean by
progressive dispersal by the currents. Ignoring the seed bank may also decrease overall
persistence, as fitting models I and II without a seed bank does not
allow all species to persist here. If the goal of a community-level
model is very short-term prediction (days, weeks), this recolonization
can probably be neglected, and there are no risks of extinction over
these short time frames. However, over multiple years, ignoring cryptic
life stages allowing recolonization could strongly bias downwards
our view of long-term coexistence. Long-term phytoplankton coexistence
modelling (over multiple decades or more) likely requires that we
take into account resting stages, whose influence may become only
more important as the timescale increases, due to the very long possible
dormancies that have been evidenced \citep{ellegaard_long-term_2018,sanyal2022not}.
When modelling different stages of the life cycle in a detailed manner
-- as done here -- is impractical, the recolonization could perhaps
be simplified to a stochastic immigration term (as done in \citealt{stock_evaluating_2005}
in a single-species context). This technical suggestion certainly
extends to models of (terrestrial) plant community dynamics.

More research on dormant stages may be needed to parameterize truly
predictive mechanistic phytoplankton models with multiple life stages,
in particular to inform parameters such as the sinking rate of resting
cells, as well as burial and resuspension parameters. These parameters
are all linked to hydrodynamics \citep{yamamoto_modelling_2002,yniguez_investigating_2012}
and may locally vary. Sinking rates show an interesting conflict between
short- and long-term survival: in coastal areas, a fraction of sinking
cells contribute to the seed bank, increasing the odds of species
long-term survival at the cost of short-term individual cell survival.
But high sinking rates are essentially ``wasted'' in the open ocean
-- whether different sinking rates can be selected, to some degree,
by such different environments could be quite revealing. How cells
get up rather than down in water column might be as interesting but
more difficult to study. The likely idiosyncratic nature of recolonization
by resting cells -- due to the contingency on local hydrodynamics
-- means that experimentation might be the only manner in which the
frequency of reinvasion can be assessed. Currently, one of the only recolonization-related
parameters measured in the field is the rate of survival of the cells
found in the sediment \citep{montresor_diversity_2013,solow_testing_2014}.
While very important, this parameter is a necessary not sufficient
condition for reinvasion of the population at future times. We need
more information about the abilities of resting cells buried in the
sediment to come up to the pelagic zone, which is required for recolonization
to actually occur. Many factors may contribute: bottom currents, benthic
animals,... We therefore encourage both experiments and field observations
to follow actual seed trajectories, in order to help us understand
this cryptic part of the diversity maintenance process.

\subsubsection*{Declaration of Competing Interest}

The authors declare that they have no known competing financial interests
or personal relationships that could have appeared to influence the
work reported in this paper.

\subsubsection*{Author contribution statement}

CP and FB designed the project and the models. CP wrote the computer code and produced
the figures. CP and FB interpreted the results and wrote the manuscript.

\subsubsection*{Acknowledgements}

We thank Nathan Wisnoski and Frank Jabot for useful feedback. FB and
CP were supported by grants ANR-10-LABX-45 and ANR-20-CE45-0004. CP
was supported by a PhD grant from the French Ministry of Research.

\bibliographystyle{ecol_let}
\bibliography{biblio_seed}

\clearpage

\textbf{Supporting Information} for \emph{Seed banks can help to maintain
the diversity of interacting phytoplankton species }by C. Picoche
\& F. Barraquand. (doi:\verb|10.1016/j.jtbi.2022.111020|)

\renewcommand\thesection{S\arabic{section}}
\setcounter{section}{0}
\renewcommand\thefigure{S\arabic{figure}}
\setcounter{figure}{0}
\renewcommand\thetable{S\arabic{table}}

\section{Intrinsic growth rate modeling}

The intrinsic growth rate can be decomposed in 2 elements:

\begin{equation}
r_{i}(T)=E(T)f_{i}(T)\label{eq:growth_rate}
\end{equation}

with $E(T)$ the response to temperature common to all species and
$f_{i}(T)$ the species-specific response. This section provides detailed
information regarding $E(T)$ and $f_{i}(T)$ estimates.

\subsection*{Common response to temperature}

Phytoplanktonic growth rates cover a broad range of values: between
0.2 and 1.78 day$^{-1}$ for diatoms in \citet{reynolds2006ecology},
even reaching 3 day$^{-1}$ in the meta-analysis of 308 experiments
by \citet{edwards_light_2015}. These values are often computed from
measurements on isolated species or on small communities in laboratory
conditions, in a constant environment. A broader perspective is therefore
necessary to understand general responses to changes in the environment
\citep{bissinger_predicting_2008,edwards_phytoplankton_2016}, especially
temperature.

\begin{figure}[H]
\begin{centering}
\includegraphics[width=0.99\textwidth]{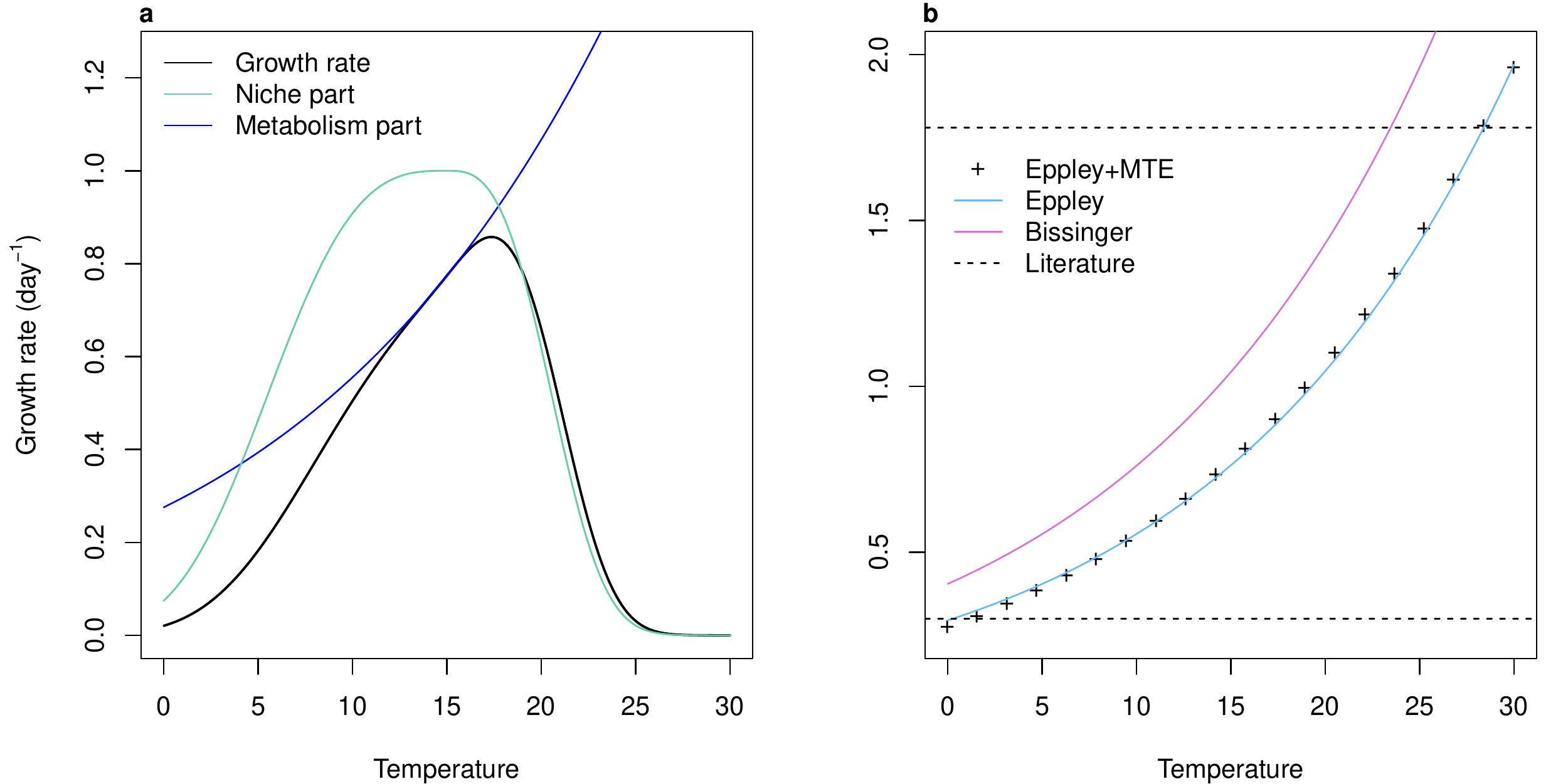}
\par\end{centering}
\caption{Decomposition of the \citet{scranton_coexistence_2016} growth rate
formula (a). The black line indicates the final growth rate with their
model. The blue line corresponds to the species-specific response
to temperature for a thermal optimum of 15°C and the green line is
the maximum achievable growth rate, a composite of the metabolic theory
of ecology (MTE) and the formula by \citet{eppley_temperature_1972}.
This formula is shown by black crosses in (b) and compared to the
\citet{eppley_temperature_1972} curve in blue and \citet{bissinger_predicting_2008}
formula in purple. Horizontal lines show limits found in the literature
\citep{reynolds2006ecology}. \label{fig:Comparison-of-growth-rate}}
\end{figure}

We first used the equation by \citet{scranton_coexistence_2016} as
a starting point, but it was not able to reproduce the observed values
found in \citet{edwards_light_2015} (see low values of the growth
rate in Fig. \ref{fig:Comparison-of-growth-rate}a, between 0.05 and
0.8, in place of values between 0.2 and 3 day$^{-1}$). In this context,
we decided to use the formula by \citet{bissinger_predicting_2008}
to compute the maximum growth rate response to the temperature. There
are two reasons for this choice. First, their model is a general function
that can be applied to all species. Second, \citet{bissinger_predicting_2008}
is an update of the seminal work of \citet{eppley_temperature_1972}
(which was used in \citealp{scranton_coexistence_2016}, but might
be outdated).

The relationship between temperature and growth rate is then $E(T)\thinspace=\thinspace0.81\exp^{0.0631(T-273.15)}$,
with $T$ in Kelvin degrees. In this case, growth rates vary between
0.81 and 3.9 day$^{-1}$ for temperatures between 0 and 25°C, in line
with previous observations. However, these daily growth rates need
to be proportional to the daylength as no growth occurs at night:
we therefore divide $E(T)$ by two in our models.

\subsection*{Species-specific response to temperature}

The niche part of the growth rate $f_{i}(T)$ is mainly defined by
two parameters which drive the phenology of a species: the thermal
optimum $T_{i}^{opt}$ and a proxy of the niche width $b_{i}$ (eq.
\ref{eq:niche_part}).

\begin{equation}
f_{i}(T)=\begin{cases}
\exp(-|T-T_{i}^{opt}|^{3}/b_{i}), & T\leq T_{i}^{opt}\\
\exp(-5|T-T_{i}^{opt}|^{3}/b_{i}), & T>T_{i}^{opt}
\end{cases}\label{eq:niche_part}
\end{equation}
The annual dynamics of phytoplanktonic organisms is usually characterized
by a blooming period and a lower concentration during the rest of
the year. The bloom can be triggered by a combination of nutrient
and light input, as well as a sufficient temperature. All these variables
being more or less correlated to the seasonal rythm, it is reasonable
to restrain this study to the effect of one variable, temperature.
Thus, the niche computed here is not the fundamental thermal niche,
but a composite of all environmental conditions covarying with the
temperature and promoting population growth. Such environmental conditions
of course include solar irradiance which is strongly correlated to
temperature in the field, but could also account for other factors,
such as predation, that have a seasonal rythm which is partly captured
by temperature. The niche part of the growth rate $f_{i}(T)$ therefore
describes the \emph{realized} niche of species $i$, not its fundamental
niche.

We base our estimates of $T_{i}^{opt}$ and $b_{i}$ on field observations.
For each taxon and each year, we define the beginning of the bloom
as the date when the taxon abundance exceeds its median abundance
over the year. The duration of the bloom is the number of days between
the beginning and the date where abundance falls below the median
value. Taxa are then separated into two groups. In the field, generalists
are characterized by one long bloom in the year or several blooms
during which the abundances oscillate around their median. Specialists
tend to appear only once or twice in the year for shorter amounts
of time. A genus is therefore defined as a generalist if the duration
of its cumulated bloom days over a year last more is above the average
duration of all blooms (137 days) for at least 15 years over the 20
years of the time series, and as a specialist if they fall below this
threshold.

In the models, we assume that generalists have a niche width between
15 and 30° and specialists, between 5 and 10°. In order to compute 
$b_{i},$ we make the additional assumption that temperatures outside
of this range lead to a growth rate at least 10 times inferior to
the growth rate obtained at their thermal optimum ($\exp(-|7.5|^{3}/b_{i}=0.1$
for a niche width of 15°). This leads to values of $b_{i}$ between
180 and 1500 for generalists, and 7 and 55 for specialists. A set
of $b$ values is drawn from a uniform distribution within these boundaries.
Meanwhile, taxa are ordered as a function of the mean cumulated bloom
length and larger niche values are attributed to longer mean bloom
length, i.e. $\sum\overline{L}_{i}>\sum\overline{L}_{j}\Rightarrow b_{i}>b_{j}$
where $\overline{L}$ is the mean over 20 years of the annual cumulated bloom
lengths.

The thermal optimum $T_{i}^{opt}$ was first defined as the mean minimum
temperature of the bloom throughout the whole time series. However,
this value led to blooms occuring mainly in the winter and needed
to be increased by 5° in order to simulate realistic phytoplankton
cycles.

\begin{figure}[H]
\begin{centering}
\includegraphics[width=0.99\textwidth]{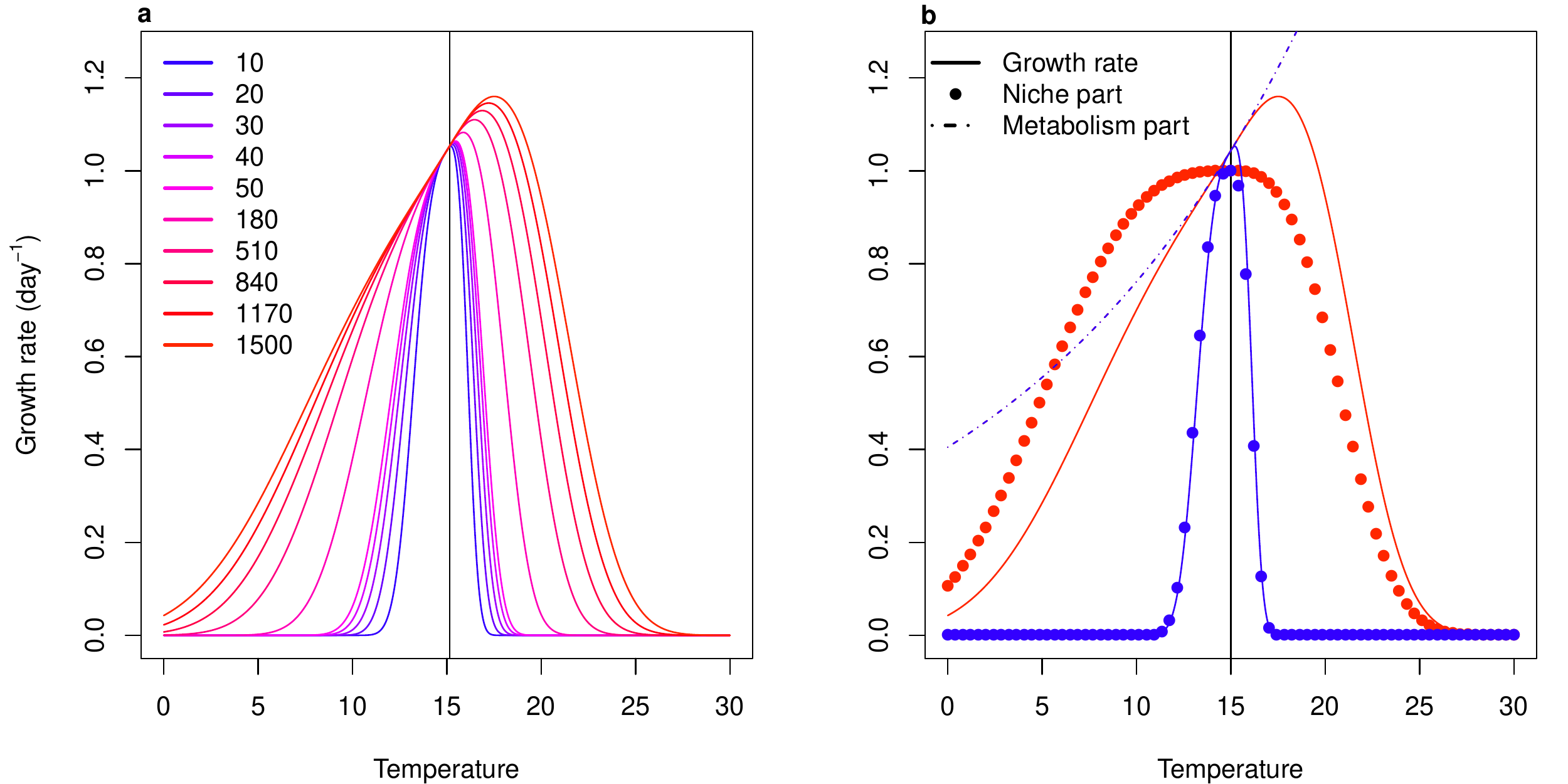}
\par\end{centering}
\caption{Relationship between daily growth rates and temperature with different
values of niche width $b$ (whose values are indicated in the legend,
corresponding to specialist and generalist species) and the same thermal
optimum, 15°C, indicated by the solid black line (a). On the right
panel, only the two extreme values of $b$ (10 and 1500) are shown
in blue and red respectively. Solid lines then correspond to the final
growth rate, points correspond to $f_{i}(T)$ values (see eq. \ref{eq:growth_rate})
and the dotted line corresponds to $E(T)$ values.\label{fig:Growth_niche_metabolism}}
\end{figure}

It should be noted that a variation in niche width also affects the
final shape of thermal preferences. Indeed, when $b_{i}$ increases,
the niche term $f_{i}$ has smaller variation in values around the
thermal optimum. In this case, the final value of the growth rate
is driven by the metabolism part of the equation (Fig. \ref{fig:Growth_niche_metabolism}).

\begin{figure}[H]
\begin{centering}
\includegraphics[width=0.95\textwidth]{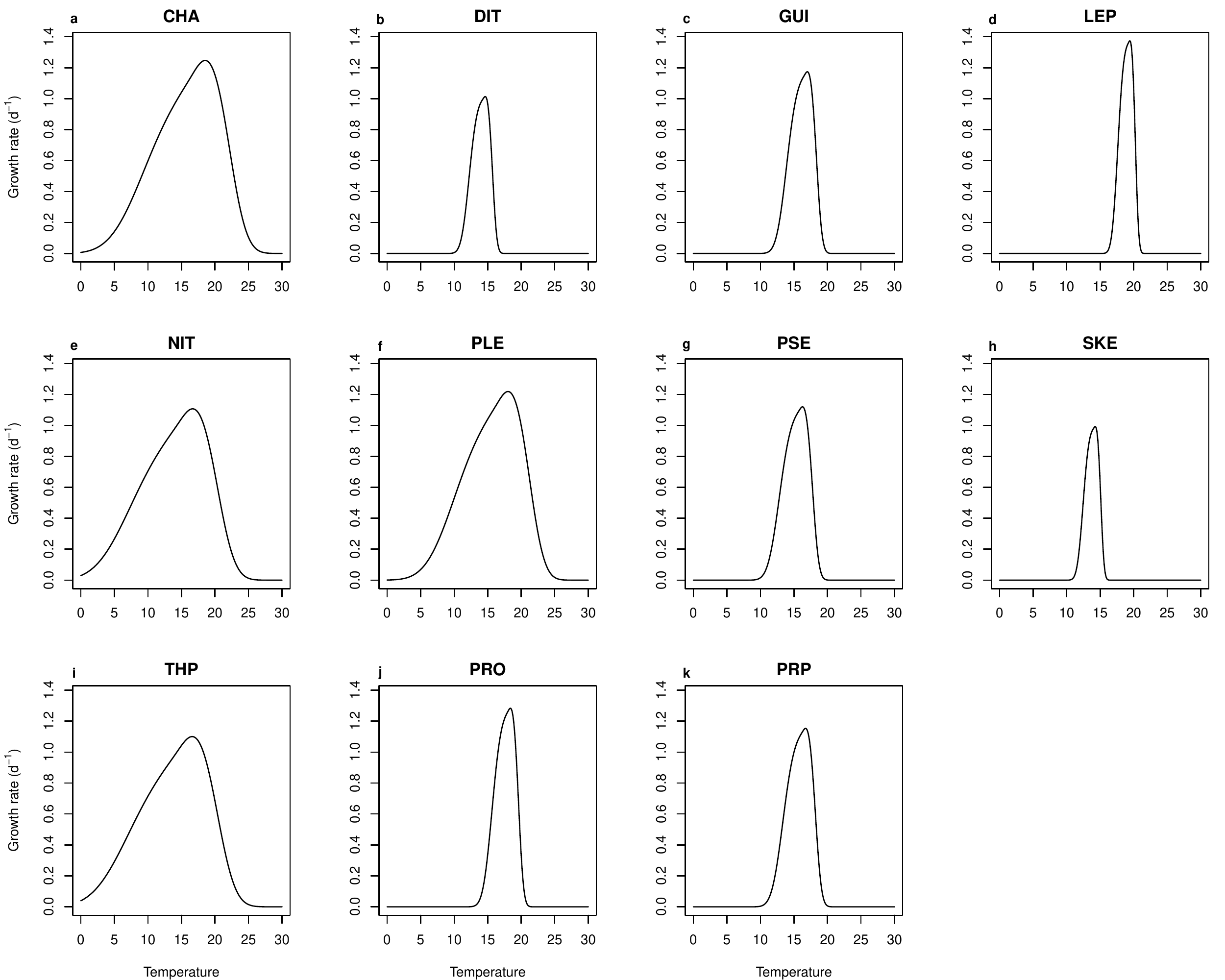}
\par\end{centering}
\caption{Growth rate as a function of temperature for the taxa used in the
model (names are defined in Table \ref{tab:Group_definition}). \label{fig:Thermal_niche}}
\end{figure}

We show the growth rates $r_{i}(T)$ as a function of temperature
for each modeled taxon below (Fig \ref{fig:Thermal_niche}).

\section{Initial estimates of interaction values}

\subsubsection*{Model I: Lotka-Volterra interactions}

Interactions between taxa have previously been computed with a Multivariate
AutoRegressive model (eq. \ref{eq:MAR_equation}, \citealp{picoche_strong_2020}).

\begin{equation}
\mathbf{n}_{t+1}=\mathbf{B}\mathbf{n}_{t}+\textbf{C}\textbf{u}_{t+1}+\mathbf{e}_{t},\mathbf{e}_{t}\thicksim\mathcal{{N_{S}}}(0,\mathbf{Q})\label{eq:MAR_equation}
\end{equation}

where $S$ is the number of taxa, $\mathbf{n}_{\ensuremath{t}}$ is
the $S\times1$ vector of log-abundance of phytoplankton taxa, $\mathbf{B}$
is the $S\times S$ interaction matrix with elements $b_{ij}$ (the
effect of taxon $j$ on taxon $i$), $\mathbf{C}$ is the $S\times V$
environment matrix describing the effects of variables $\mathbf{u}_{t+1}$
on growth rates and the noise $\mathbf{e}_{t}$ is a $S\times1$ noise
vector following a multivariate normal distribution with a variance-covariance
matrix $\mathbf{Q}$. The interaction model we use in the present
paper is a Beverton-Holt multispecies model (eq. 1 in main text),
also called at times Leslie-Gower. In \citet{picoche_strong_2020}'s
Supporting Information, we showed that MAR and BH interaction coefficients,
respectively $b_{ij}$ and $\alpha_{ij}$, could map once abundances
at equilibrium $N_{i}^{*}$ are defined.

\[
\begin{cases}
b_{ii}-1= & \frac{-\alpha_{ii}N_{i}^{*}}{1+\sum_{l}\alpha_{il}N_{l}^{*}}\\
b_{ij,i\neq j}= & \frac{-\alpha_{ij}N_{j}^{*}}{1+\sum_{l}\alpha_{il}N_{l}^{*}}
\end{cases}
\]

Let's define $\tilde{b}_{ij}$ with $\tilde{b}_{ii}=b_{ii}-1$, and
$f_{A}(i)=\sum_{l}\alpha_{ij}N_{l}^{*}$.

\[
\tilde{b}_{ij}(1+f_{A}(i))=-\alpha_{ij}N_{j}^{*}
\]

We then sum on columns (on $j$):

\[
\sum_{j}[\tilde{b}_{ij}(1+f_{A}(i))]=-f_{A}(i)
\]

\[
\Leftrightarrow-f_{A}(i)(1+\sum_{j}\tilde{b}_{ij})=\sum_{j}\tilde{b}_{ij}
\]

\[
\Leftrightarrow f_{A}(i)=-\frac{\sum_{j}\tilde{b}_{ij}}{(1+\sum_{j}\tilde{b}_{ij})}
\]

\[
\Leftrightarrow\alpha_{ij}=-\frac{1}{N_{j}^{*}}\tilde{b}_{ij}(1-\frac{\sum_{j}\tilde{b}_{ij}}{1+\sum_{j}\tilde{b}_{ij}})
\]

\[
\Leftrightarrow\alpha_{ij}=-\frac{1}{N_{j}^{*}}\frac{\tilde{b}_{ij}}{1+\sum_{j}\tilde{b}_{ij}}.
\]

This gives an exact correspondence between $\alpha_{ij}$ and $b_{ij}$.
In the multispecies BH model, the presence of mutualistic interactions
can lead to an orgy of mutual benefaction \citep{may_1981_theoretical}.
We impose a minimum value of 1 to the denominator of the BH formulation,
meaning that the growth rate cannot be higher than the maximum growth
rate calculated, $r_{i}(T)$. 

\subsubsection*{Model II: saturating interactions}

We now move to a model with saturating interactions between taxa:
\begin{equation}
N_{t+1,i}=\frac{e^{r_{i}(T)}N_{t,i}}{1+\sum_{j/a\in\mathbb{C}}\frac{a_{C}N_{t,j}}{H_{ij}+N_{t,j}}+\sum_{j/a\in\mathbb{F}}\frac{a_{F}N_{t,j}}{H_{ij}+N_{t,j}}}\label{eq:saturation_h}
\end{equation}

where coefficients $a_{C}$ and $a_{F}$ are the maximum interaction
strengths for competition and facilitation, respectively, $H_{ij}$
is the abundance of taxon $j$ to reach half of the maximum effect
of taxon $j$ on taxon $i$, and $\mathbb{C}$ and $\mathbb{F}$ are
the sets of competitive and facilitative interactions. This formula
can be linked to the Unique Interaction Model by \citet{qian_balance_2020},
i.e., each taxon provides a unique type of benefit or disadvantage
to the focus taxon.

There is no single solution for matching the $\mathbf{B}$ matrix
of the MAR model to model II including $H_{ij}$, $a_{C}$ and $a_{F}$.
We approximate the maximum interaction strength $a_{C}$ as the average
sum of all taxon effects $\alpha_{ij}N_{j}$ exerted on a given taxon
if all interactions were competitive (eq. \ref{eq:a_C}). To compute
$a_{F}$, we make two assumptions: on average, a) there is 70\%
facilitation in our dataset and b) the realized growth rate on a log-scale should
not exceed $r_{i}(T)$, as in model I. We consider that the relationships that
apply to individual interactions $\alpha_{ij}$ should also apply
to the saturation point (eq. \ref{eq:a_F}), so that:
\begin{eqnarray}
a_{C} & = & \frac{1}{S}\sum_{i}\left(\sum_{j}|\alpha_{ij}|N_{j,\text{max}}\right)\label{eq:a_C}\\
(1-0.7)a_{C}+0.7a_{F} & = & 0\label{eq:a_F}
\end{eqnarray}

where $N_{j,\text{max}}$ is the maximum observed abundance of species $j$.
We use $N_{j,\text{max}}$ and the absolute value of interactions $|\alpha_{ij}|$
(i.e., all interactions are considered competitive in this case) to
make sure that we maximise $a_{C}$.

At low abundances, we can consider that interactions are far from
saturation. Taking the tangent of the function at this point, $H_{ij}$
can be approximated by $f\frac{a_{C/F}}{\alpha_{ij}}$, where $f=2$
is a correction factor that takes into account the fact that the slope
at origin for the type II response is likely higher than the slope
for a linear effect of density.

\section{Choice of parameters derived from literature}

This section contains additional information on ``fixed'' parameter
definitions (i.e., parameter not estimated from field data) and their
chosen values. Note that these values are then subjected to sensitivity
analysis, where by definition they are modified.

\paragraph*{Loss rate}

The loss rate corresponds to multiple mortality processes. The Lotka-Volterra
model of \citet{scranton_coexistence_2016} considered a rate around
0.04 day$^{-1}$. In \citet{jewson_loss_1981}, washout (0.5\%), parasitism
(4\% of cells are infested and die) and grazing still remained low
(about 0.05\%) when compared to growth rates. \citet{li_what_2000}
found values between 0.02 and 0.1 day$^{-1}$ for natural mortality
only, while a review by \citet{sarthou_growth_2005} indicated a loss
of daily primary productivity around 45\% due to grazing only, while
cell autolysis only can lead to a loss rate between 0.005 and 0.24
day$^{-1}$ (in the absence of nutrients, or because of viral charge).
Cumulating both natural mortality (cell autolysis) and grazing, we
know that maximum rates should be above 0.24 day$^{-1}$. Trying to
make a compromise emerge from the literature above, given that we
are not always at the maximal loss rates, we set the reference value
to 0.2 day$^{-1}$. This strikes a balance between acknowledging all
sources of mortality and the need to keep all species in the reference
model.

\paragraph*{Sinking rate}

\begin{figure}[H]
\begin{centering}
\includegraphics[width=0.5\textwidth]{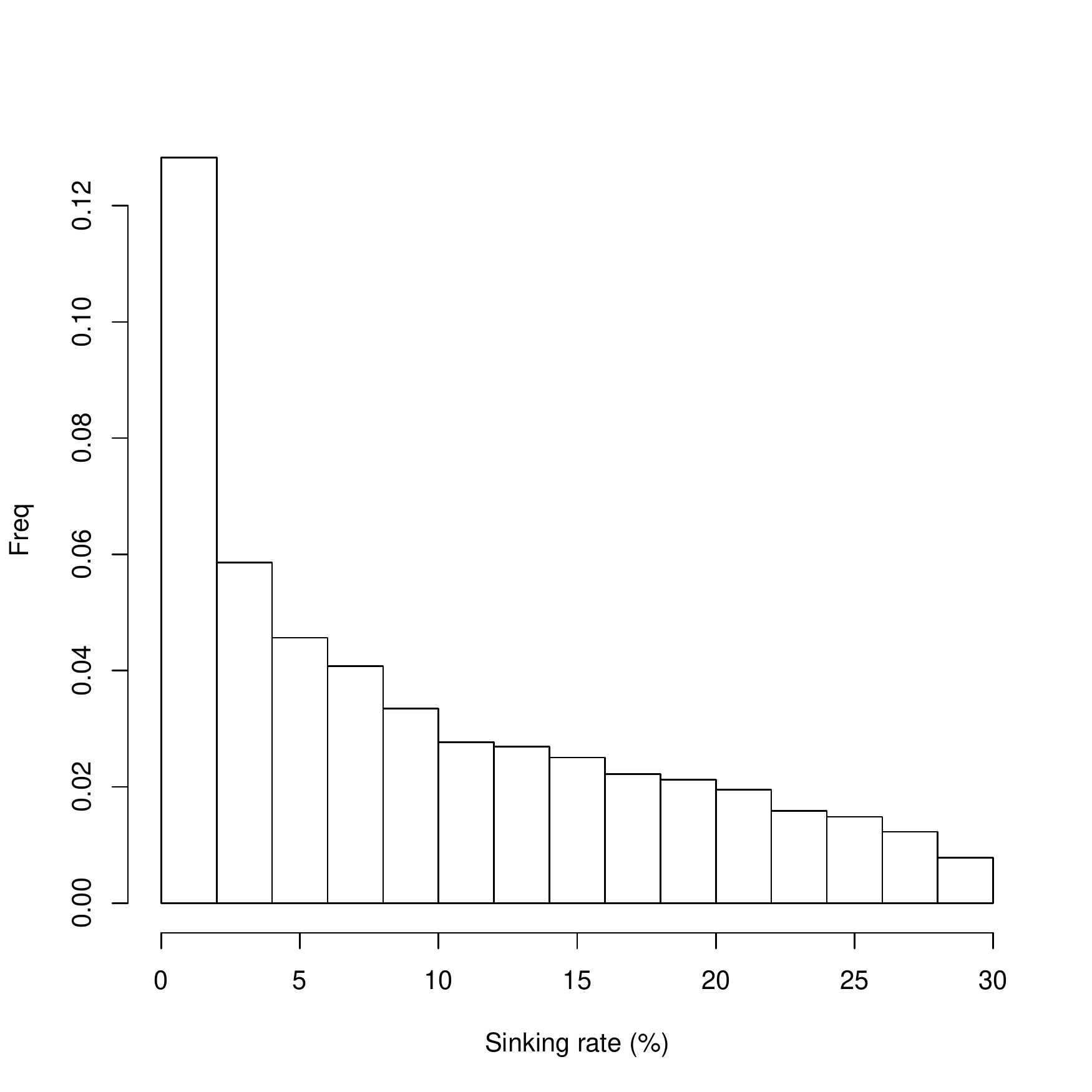}
\par\end{centering}
\caption{Assumed Beta-distribution of sinking rates\label{fig:Sinking-distribution}}
\end{figure}

Among the hydrodynamics processes that drive the sinking rate, turbulence
and eddies -- themselves driven by tidal currents, the shape of the
coast or wind conditions -- are influential in keeping the cells
at the top of the water column. For that reason, laboratory experiments
on sinking rates are not sufficient to calibrate a field-based model.
We therefore chose sinking rate values from field studies. In the
Gotland Basin (central Baltic Sea), \citet{passow_species-specific_1991}
measured a large variability in sinking rates, even within the same
genus (e.g., between 1 and 30\% for \textit{Chaetoceros} spp.). However,
a pattern could be highlighted, with a small number of genera that
sank more than the rest of the community. The mean sinking rate for
\textit{Chaetoceros} and \textit{Thalassiosira} was around 10\% while
it was around 1\% for the other species. Sinking rate values around
10\% are consistent with the loss rates in \citet{kowe_modelling_1998}
in a river and \citet{wiedmann_seasonality_2016} in an estuary (mouth
of Adventfjorden). When estimating changes of the sinking rate over
time, values between 4 and 50\% were obtained \citep{jewson_loss_1981}.
We therefore chose to represent the sinking rates with a Beta-distribution
(Fig. \ref{fig:Sinking-distribution}) which accounts for observed
maximum and mean values, while still allowing a highly skewed distribution
of sinking rates between species. High sinking rates are attributed
to the morphotypes corresponding to \textit{Chaetoceros} (CHA) and
\textit{Thalassiosira }(THP)\textit{.}

\paragraph*{Mortality and burial in the seed bank}

\citet{mcquoid_viability_2002} present maximum and mean depth of
sediment at which germination of diatoms and dinoflagellates could
still occur when incubated. The authors also present sediment datation
according to depth. Depth can therefore be related to maximum and
mean age of phytoplankton resting cells before death. Assuming $m$
is the probability of mortality and survival follows a geometric law,
the life expectancy of a resting cell is $\frac{\ensuremath{1}}{m}\Leftrightarrow m=\frac{1}{L_{\text{mean}}}$
where $L_{\text{mean}}$ is the average duration of the dormant cell viability.
Another way to look at the process is that life expectancy $L$ follows
the distribution $p(L>l)=e^{-ml}.$ We arbitrarily chose that for
the oldest dormant cells (i.e., the ones buried the deepest), $p(L>l_{\text{max}})=0.05$.
Following this, $m=-\frac{\ln(0.05)}{l_{\text{max}}}$ where $l_{\text{max}}$
derives from maximum depth values (for each species) in \citet{mcquoid_viability_2002}.
For both methods, $m$ varies between $10^{-5}\text{ \text{day}}^{-1}$
and $10^{-4}\text{\text{ day}}^{-1}$ for all species considered.

As we highlight in the main text, burial is a very important process
that controls the availability of resting cells, conditional to their
survival in the sediment. However, burial rate is almost entirely
dependent on the local sedimentation and no generally applicable literature
could be found.  We varied the burial rate $\zeta$ between 0.001
and 0.1 per day.

In scenarios where we remove the seed bank, we set $m+\zeta$ to 100\%
(for simplicity, we do not eliminate resting stage formation, only resting
stage survival). 

\paragraph*{Resuspension}

As mentioned in the main text, resuspension values are mostly taken
from models or data for inorganic particles. Rates vary greatly from
one publication to another: in \citet{fransz_modelling_1985}, in
a coastal area, the resuspension rate of sediments is evaluated around
$5\times10^{-5}$ day$^{-1}$ in winter and decreases in summer, with
a relationship between resuspension and the light extinction coefficient.
In \citet{kowe_modelling_1998}, the resuspension rate of diatoms
is evaluated around $1.9\times10^{-5}$ day$^{-1}$. In \citet{le_pape_pelagic_1999},
resuspension rate of sediments and dead diatoms is 0.002 day$^{-1}$.
In this paper, we explore values between $10^{-5}$ (stratified water
column) to 0.1 (highly mixed environment).

Finally, it should be noted that burial, sinking rate and resuspension
are all highly contingent upon the local hydrodynamics and therefore
are intermingled processes.

\section{Phytoplankton taxa at the calibration site}

\begin{table}[H]
\begin{centering}
\begin{tabular}{cc}
\hline 
\textbf{Code} & \textbf{Taxa}\tabularnewline
\hline 
CHA & \textit{Chaetoceros}\tabularnewline
DIT & \textit{Ditylum}\tabularnewline
GUI & \textit{Guinardia}\tabularnewline
LEP & \textit{Leptocylindrus}\tabularnewline
NIT & \textit{Nitzschia+Hantzschia}\tabularnewline
PLE & \textit{Pleurosigma+Gyrosigma}\tabularnewline
PRO & \textit{Prorocentrum}\tabularnewline
PRP & \textit{Protoperidinium+Archaeperidinium+Peridinium}\tabularnewline
PSE & \textit{Pseudo-nitzschia}\tabularnewline
SKE & \textit{Skeletonema}\tabularnewline
THP & \textit{Thalassiosira+Porosira}\tabularnewline
\end{tabular}
\par\end{centering}
\caption{Name and composition of the phytoplanktonic groups used in main text\label{tab:Group_definition},
see \citet{picoche_strong_2020} for more information on these taxa.}
\end{table}

\section{Phytoplankton time series with model II}

\begin{figure}[H]
\begin{centering}
\includegraphics[height=0.8\textheight]{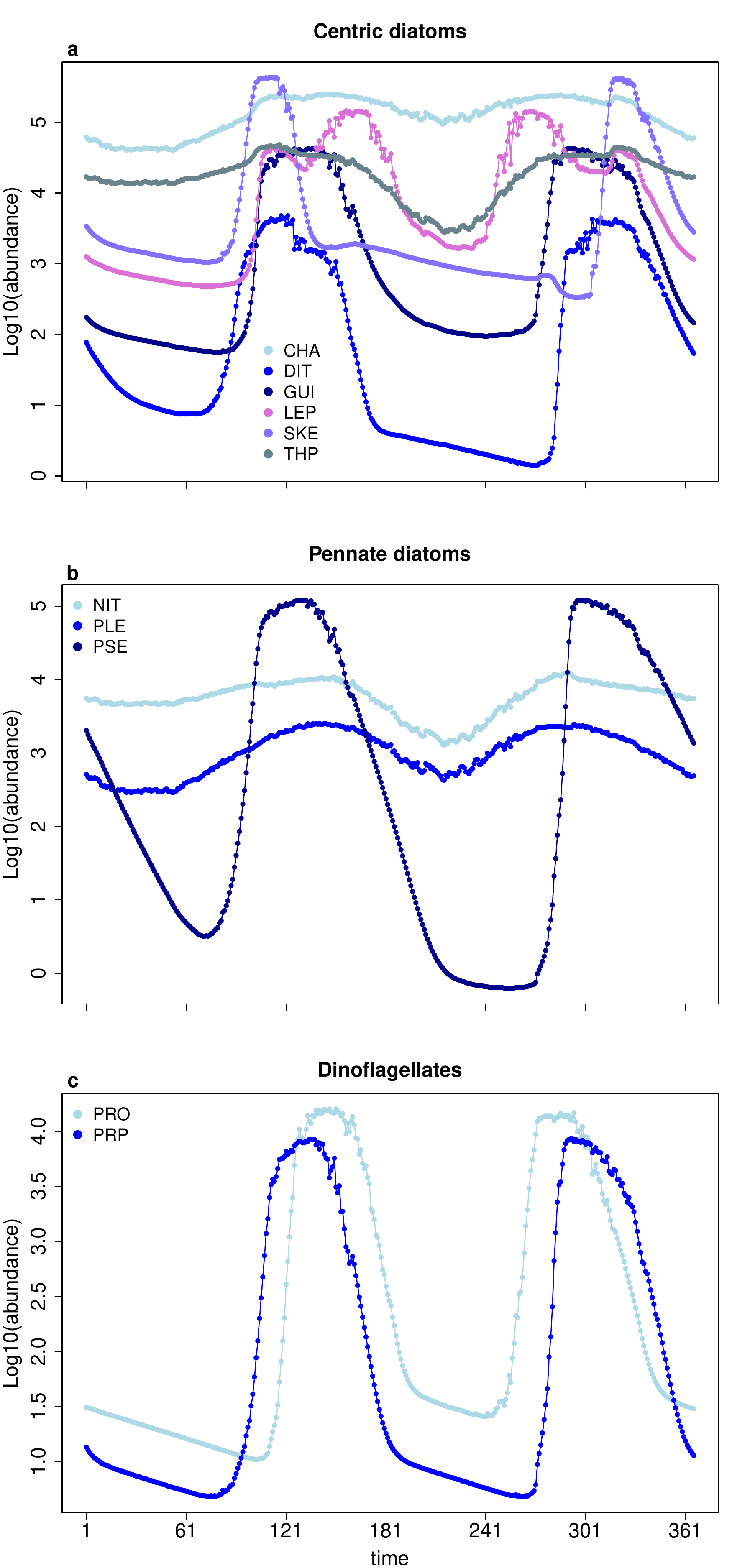}
\par\end{centering}
\caption{Simulated phytoplankton time series for a year in model II (with saturating
interactions). Each panel corresponds to a cluster of interacting
taxa: centric diatoms (a), pennate diatoms (b) and dinoflagellates
(c). Taxa only interact within their cluster (see Methods in main
text).}
\end{figure}

\section{Scenario: changing interspecific interaction strengths but not intraspecific
interaction strengths}

\begin{figure}[H]
\begin{centering}
\includegraphics[width=0.74\textwidth]{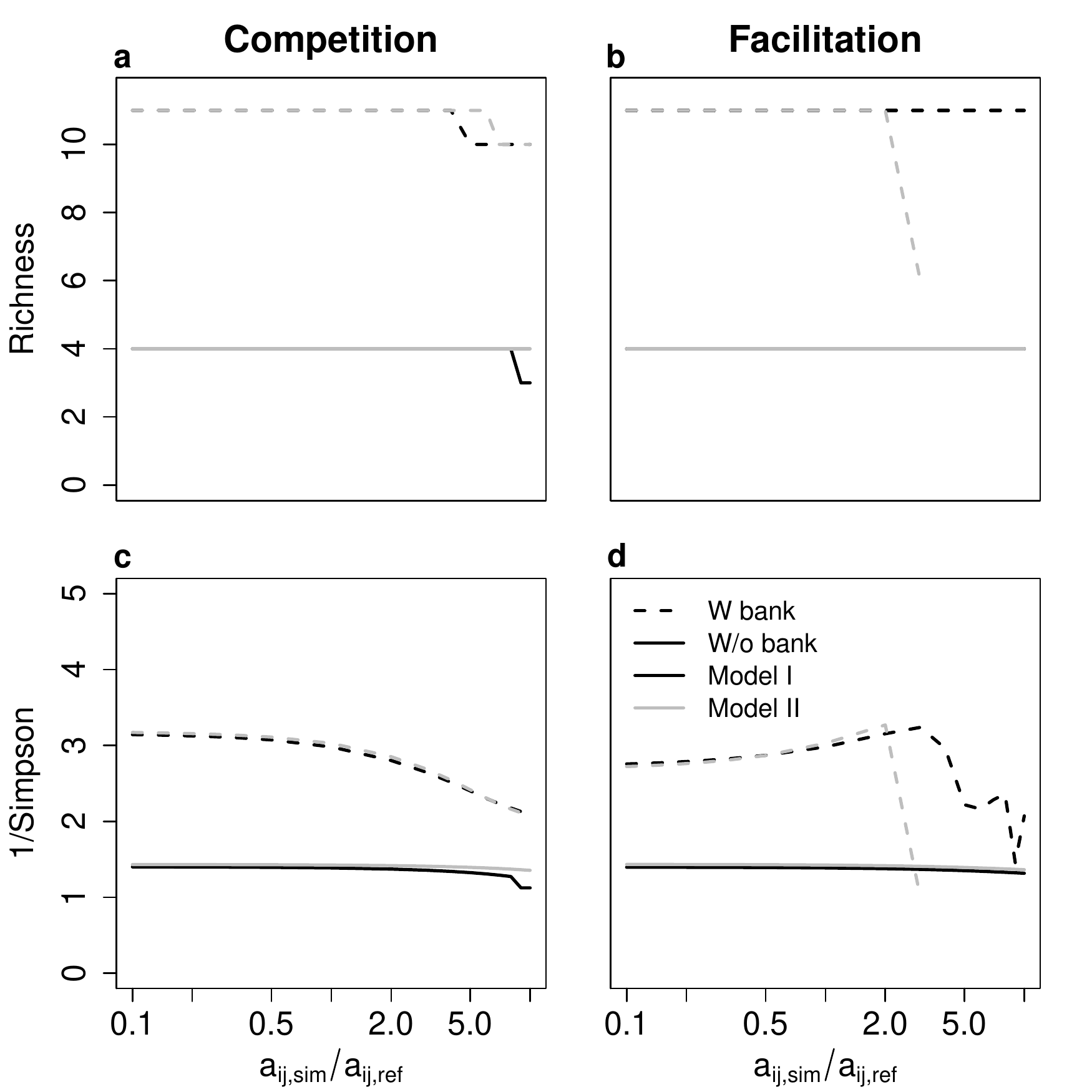}
\par\end{centering}
\caption{Measures of biodiversity in the ocean at the end of the simulation:
a-b) richness and c-d) inverse of the Simpson index, with (dashed
line) and without (solid line) a seed bank, as a function of the strength
of competition and facilitation with a classical Beverton-Holt (black
lines) or a saturating interaction (grey lines) formulation, keeping
the same values for the intraspecific interaction strengths. The x-axis
shows the factor by which each interaction coefficient was multiplied,
e.g. the value 0.1 indicates that the interaction strengths in the
simulation are 10 times lower than the interactions strengths in the
reference simulation. Note the logarithmic scale. \label{fig:Seed_bank_compet}}
\end{figure}

When intraspecific interaction strengths do not change, reducing the
values of interspecific competition has very little effect on both
richness and the inverse of the Simpson index. Increasing facilitation
finally destabilizes the community and leads to the observed
diversity decrease in the absence of a seed bank. We do stress, however,
that this scenario is more of a thought experiment changing niche
differentiation rather than anything mimicking an environmental perturbation,
where the factors that change the degree of competition between species
will likely change competition within species too.

\section{Growth rates and survival without a seed bank}

In order to investigate the relationships between population dynamics
and survival probabilities, we computed the realized per capita growth
rates (PCGR$_i=\exp(r_i(T))/C_i$, where the competition $C_i$ is defined in Section S8) of each species in simplified conditions (Fig. \ref{fig:Probability-of-survival}):
\begin{itemize}
\item when all species abundances are set to 1, so that there is nearly
no competition
\item when all species abundances are set to their average abundance in
the environment, as a proxy of competition endured by each species
throughout the year
\item when temperature is set to the average temperature of the environment
\item when temperature is either optimal for each species, or too high for
all species
\end{itemize}
The population grows when PCGR$_i>1+l$, with $l$ the loss rate.

It should be noted that the realized growth rate is not computed using
long-term simulation, but for a set of fixed environmental and competition
values, in the abovementioned conditions. Growth rates are then related
to the probability of survival in the ocean without a seed bank when
varying interaction strengths (scenario 1 in the main text). Probability
of survival is itself computed over different values of interaction
strengths.

\begin{figure}[H]
\begin{centering}
\includegraphics[width=0.75\textwidth]{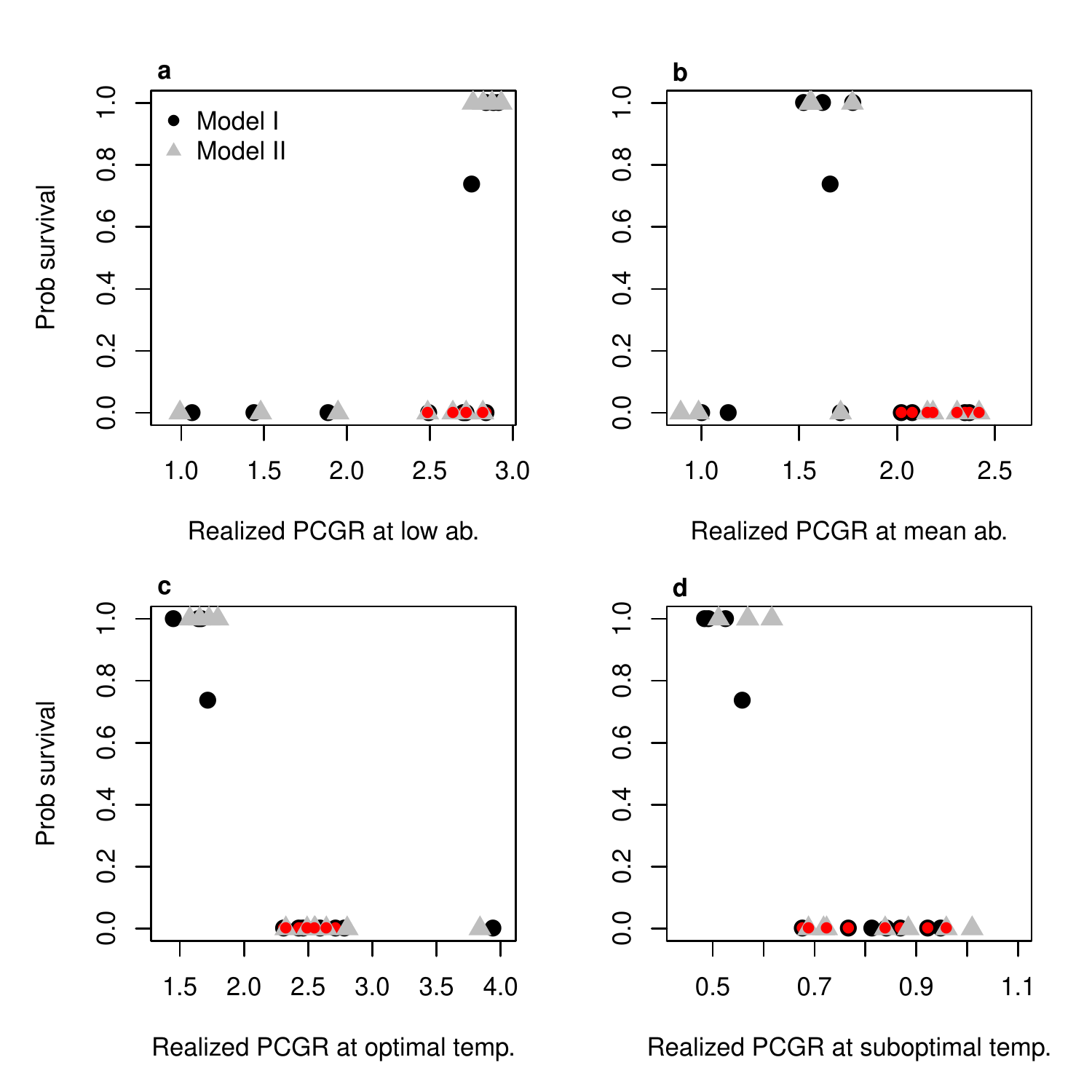}
\par\end{centering}
\caption{Probability of survival in the absence of a seed bank, as a function
of the realized per capita growth rates (PCGR) of each species in
different conditions. We consider (a) mean temperature and low densities,
i.e., temperature is set to 15°C and all species densities are 1 cell/L,
or (b) mean temperature (15°C) and average densities, i.e., all species
are at their simulated average density values; or (c) optimal temperature
for each species and average densities, or (d) suboptimal temperatures
(30°C) and average densities. Red points correspond to species which
have a high growth rate at low density but still go extinct. \label{fig:Probability-of-survival}}
\end{figure}

Extinction in the absence of the seed bank is mostly due to a narrow niche: extinct species are always specialists (Fig. 5), even though these species should be able to invade a typical environment (Fig. \ref{fig:Probability-of-survival} a) and tend to be less regulated than surviving species (Fig. \ref{fig:Probability-of-survival} c, d).

\section{Absence of a storage effect in the model}

The first step of our model (eq. 1 and 3 in main text) describes the
increase in abundance of coastal and oceanic populations due to both
environmental fluctuations and interactions with other organisms.
The formula may be interpreted by certain readers as already including
a storage effect, as shown by \citet{miller_evolutionary_2017} for a similar continuous-time
model, or due to its apparent similarity to the model of \citet{chesson_quantifying_2003}
in discrete time. We show below why the results of the cited analyses
do not apply here and that the intra-compartment growth in discrete-time
model does not lead to a storage effect.

The storage effect requires three elements: (a) a positive covariance
between good environmental conditions and competition, (b) species-specific
environmental responses and (c) subadditivity of environmental and
competitive effects on the growth rate \citep{chesson_quantifying_2003}.
Condition (a) is met as covariance may be created by temporally correlated
fluctuations in the environmental signal \citep{schreiber_positively_2021}.
Condition (b) is also met, as shown in Figure S3. Condition (c), however,
is not met. Subadditivity of environmental and competitive effects
can be mathematically expressed as $\frac{\partial}{\partial E}\left(\frac{\partial g}{\partial C}\right)<0$
where $g$ is the growth rate, and $E$ and $C$ are the effects of
the environment and competition on the growth rate respectively. 

In model I, the growth rate of a population is defined by the following
equation:

\[
N_{t+1,i}=\frac{\exp(r_{i}(T))N_{t,i}}{1+\sum_{j}\alpha_{ij}N_{t,j}}-lN_{t,i}
\]

Here $E_{i}(t)=\exp(r_{i}(T))$ and $C_{i}(t)=1+\sum_{j}\alpha_{ij}N_{t,j}$.
Therefore,
\[
\begin{array}{ccc}
g_{i} & = & \log\left(\frac{N_{t+1,i}}{N_{i}}\right)\\
 & = & \log\left(\frac{\exp(r_{i}(T))}{1+\sum_{j}\alpha_{ij}N_{t,j}}-l\right)\\
 & = & \log\left(\frac{E}{C}-l\right)
\end{array}
\]

\citet{miller_evolutionary_2017} consider instead a per capita growth rate (fitness) in continuous time $\frac{1}{N_{i}}\frac{dN_{i}}{dt}$.

While the discrete-time formula is at first sight very similar to
the one in \citet{chesson_quantifying_2003}, we wish to draw the
reader's attention to the fact that, contrary to \citet{chesson_quantifying_2003}, interaction
strengths do not depend on environmental effects, nor are they linearly
correlated.

If we take derivatives with regards to competition effect $C$:

\[
\begin{array}{ccc}
\frac{\partial g}{\partial C} & = & -\frac{1}{C^{2}}\frac{E}{\frac{E}{C}-l}\\
 & = & -\frac{E}{CE-lC}
\end{array}
\]

Finally, we take a derivative with respect to $E$:

\[
\begin{array}{ccc}
\frac{\partial}{\partial E}\left(\frac{\partial g}{\partial C}\right) & = & -\frac{CE-lC-EC}{(CE-lC)^{2}}\\
 & = & \frac{lC}{(CE-lC)^{2}}
\end{array}
\]

As $C=1+\sum_{j}\alpha_{ij}N_{t,j}$ in
the first model, or $C=1+\sum_{j\in\mathbb{C}}\frac{a_{C}N_{t,j,c/o}}{H_{ij}+N_{t,j,c/o}}+\sum_{j\in\mathbb{F}}\frac{a_{F}N_{t,j,c/o}}{H_{ij}+N_{t,j,c/o}}$
in the second model, which is never negative and always above 1, we always have $\frac{\partial}{\partial E}\left(\frac{\partial g}{\partial C}\right)>0$.
There is no buffered growth in the effective growth rate itself, and
therefore no storage effect in the coastal or oceanic compartment dynamics.

Now we highlight a more delicate point, in models that include the
seed bank compartment. The storage effect could also be due to the
presence of the seed bank, whose exchanges with other compartments
are described in the second step of the model (eq. 4 in main text).
The presence of a seed bank is indeed often associated with a storage
effect maintaining coexistence \citep{angert_functional_2009}, even
though models show that this association is not systematic \citep{aikio_seed_2002}.
However, for the storage effect to happen due to the seed bank, theory
currently highlights that environmental variations have to affect
the recruitment rate, i.e. in our case, the germination and resuspension
rates \citep{chesson_quantifying_2003}. In our model, neither environmental
nor competition effects affect the exchanges with the seed bank: the
sinking and germination/resuspension rates are fixed and independent
from both past and current environmental conditions. The storage effect
is therefore unlikely to happen in our model. A thorough analysis
of the storage effect based on simulations, as described by \citet{ellner_how_2016},
would however be needed to conclude with complete certainty, as the
presence of three distinct compartments may create non-obvious non-addivities
or non-linearities. 

\end{document}